\documentclass[12pt]{article}
\usepackage{graphics}
\usepackage{amssymb}
\usepackage{amsmath}

\usepackage{bm}

\newcommand{\rf}[1]{(\ref{#1})}
\newcommand{\beq}{\begin{equation}}

\newcommand{\eeq}{\end{equation}}
\newcommand{\bea}{\begin{eqnarray}}
\newcommand{\eea}{\end{eqnarray}}

\newcommand{\e}{\mbox{e}}
\renewcommand{\d}{\mbox{d}}

\renewcommand{\l}{\lambda}
\newcommand{\La}{\Lambda}

\renewcommand{\a}{\alpha}
\newcommand{\n}{\nu}

\newcommand{\om}{\omega}
\newcommand{\del}{\delta}
\newcommand{\Del}{\Delta}

\renewcommand{\k}{\kappa}

\newcommand{\oh}{\frac{1}{2}}

\newcommand{\tr}{\mathrm{tr}\,}
\newcommand{\ra}{\rangle}
\newcommand{\la}{\langle}

\newcommand{\mi}{\!-\!}
\newcommand{\equ}{\!=\!}
\newcommand{\plu}{\!+\!}

\newcommand{\cD}{{\cal D}}

\newcommand{\cT}{{\cal T}}

\newcommand{\cO}{{\cal O}}

\newcommand{\ts}{{\tilde{s}}}
\newcommand{\tC}{{\tilde{C}}}

\newcommand{\tk}{{\tilde{k}}}

\newcommand{\hH}{{\hat{H}}}

\newcommand{\hA}{{\hat{A}}}
\newcommand{\hC}{{\hat{C}}}
\newcommand{\hX}{{\hat{X}}}
\newcommand{\hP}{{\hat{P}}}
\newcommand{\hY}{{\hat{Y}}}
\newcommand{\hI}{{\hat{I}}}
\newcommand{\hD}{{\hat{D}}}
\newcommand{\hPk}{\hP^{kin}}
\newcommand{\hPp}{\hP^{pot}}

\newcommand{\bN}{{\bar{N}}}

\begin{document}

{\normalsize \hfill ITP-UU-08/44}\\
\vspace{-1.5cm}
{\normalsize \hfill SPIN-08/35}\\
${}$\\

\begin{center}
\vspace{24pt}
{ \large \bf The Nonperturbative Quantum de Sitter Universe}

\vspace{30pt}

{\sl J. Ambj\o rn}$\,^{a,c}$,
{\sl A. G\"{o}rlich}$\,^{b}$,
{\sl J. Jurkiewicz}$\,^{b}$
and {\sl R. Loll}$\,^{c}$

\vspace{24pt}
{\footnotesize

$^a$~The Niels Bohr Institute, Copenhagen University\\
Blegdamsvej 17, DK-2100 Copenhagen \O , Denmark.\\
{ email: ambjorn@nbi.dk}\\

\vspace{10pt}

$^b$~Institute of Physics, Jagellonian University,\\
Reymonta 4, PL 30-059 Krakow, Poland.\\
{ email: atg@th.if.uj.edu.pl, jurkiewicz@th.if.uj.edu.pl}\\

\vspace{10pt}

$^c$~Institute for Theoretical Physics, Utrecht University, \\
Leuvenlaan 4, NL-3584 CE Utrecht, The Netherlands.\\
{ email: r.loll@phys.uu.nl}\\

\vspace{10pt}
}
\vspace{48pt}

\end{center}


\begin{center}
{\bf Abstract}
\end{center}
The dynamical generation of a four-dimensional classical universe from nothing but 
fundamental quantum excitations at the Planck scale is a long-standing
challenge to theoretical physicists. 
A candidate theory of quantum gravity which achieves this goal without
invoking exotic ingredients or excessive fine-tuning 
is based on the nonperturbative and background-independent technique of Causal 
Dynamical 
Triangulations. We demonstrate in detail how in this approach a macroscopic 
de Sitter universe, accompanied by small quantum fluctuations,
emerges from the full gravitational path integral, 
and how the effective action determining its dynamics can be reconstructed 
uniquely from Monte Carlo data. 
We also provide evidence that it may be
possible to penetrate to the sub-Planckian regime, where the 
Planck length is large compared to the lattice spacing of the underlying
regularization of geometry.
\newpage

\section{Introduction}\label{intro}

A major unsolved problem in theoretical physics is to reconcile
the classical theory of general relativity with quantum mechanics.
Of the numerous attempts, some have postulated new and
so far unobserved ingredients, while others have proposed radically 
new principles governing physics at the as yet untested Planckian energy scale. 
Here we report on a much more mundane approach using only standard quantum 
field theory. In a sum-over-histories approach we will attempt to 
define a nonperturbative quantum field theory which has as
its infrared limit ordinary classical general relativity and at the same time has a nontrivial 
ultra\-vio\-let limit. From this point of view it is in the 
spirit of the renormalization group approach, first advocated 
long ago by Weinberg \cite{weinberg}, 
and more recently substantiated by several groups of researchers 
\cite{reuteretc}.
However, it has some advantages compared to the renormalization group
approach in that it allows
us to study (numerically) certain geometric observables which are difficult
to handle analytically.

We define the path integral of quantum gravity nonperturbatively using
the lattice approach known as  
{\it causal dynamical triangulations} (CDT) as a regularization.
In Sec.\ \ref{cdt} we give a short description of the formalism, providing
the definitions which are needed later to describe the measurements.
CDT establishes a nonperturbative way of performing the sum over 
four-geometries (for more extensive definitions, see \cite{ajl4d,blp}).
It sums over the class of piecewise linear four-geometries which can be 
assembled from four-dimensional simplicial building blocks of link length $a$, 
such that only {\it causal} spacetime histories are included.
The continuum limit of such a lattice theory should ideally be obtained
as for QCD defined on an ordinary fixed lattice, where for an observable $\cO(x_n)$,
$x_n$ denoting a lattice point, one can measure the correlation length $\xi(g_0)$ from
\beq\label{1.1}
-\log (\la  \cO(x_n) \cO(y_m) \ra) \sim |n-m|/ \xi(g_0) + o(|n-m|).
\eeq
A continuum limit of the lattice theory may then exist if it is possible to
fine-tune the bare coupling constant $g_0$ of the theory to a critical value
$g_0^c$ such that
the correlation length goes to infinity, $\xi(g_0) \to \infty$. Knowing how $\xi(g_0)$ 
diverges for $g_0 \to g_0^c$ determines how the lattice 
spacing $a$ should be taken to zero as a function of the coupling constants, namely
\beq\label{1.2}
\xi(g_0) = \frac{1}{|g_0-g_0^c|^\n},~~~~~a(g_0) = |g_0-g_0^c|^\n.
\eeq

The challenge when searching for a {\it field theory} 
of quantum gravity is to find a theory
which behaves in this way. The challenge is three-fold: (i) to find a suitable
nonperturbative formulation of such a theory which satisfies a 
minimum of reasonable requirements, (ii) to find
observables which can be used to test relations like \rf{1.1}, and (iii) to
show that one can adjust the coupling constants of the
theory such that \rf{1.2} is satisfied. Although we will focus on (i) in what
follows, let us immediately mention that (ii) is notoriously difficult in a theory of 
quantum gravity, where one is faced with a number of questions originating in the
dynamical nature of geometry.
What is the meaning of distance when integrating
over all geometries? How do we attach a meaning to local spacetime
points like $x_n$ and $y_n$? How can
we define at all local, diffeomorphism-invariant quantities in the continuum
which can then be translated to the regularized (lattice) theory? --
What we want to point out here is that although (i)-(iii) are
standard requirements when relating critical phenomena
and (Euclidean) quantum field theory, gravity {\it is} special and 
may require a reformulation of (part of) the standard scenario
sketched above. We will return to this issue when we discuss our
results in Sec.\ \ref{discussion}.

Our proposed nonperturbative formulation
of four-dimensional quantum gravity has a number of nice features.
Firstly, it sums over a class of piecewise linear geometries, which -- as usual -- are
described without the use of coordinate systems.
In this way we perform the sum over 
geometries directly, avoiding the cumbersome procedure of first introducing 
a coordinate system and then getting rid of the ensuing gauge redundancy,
as one has to do in a continuum calculation. 
Our underlying assumptions are that 1) the class of piecewise
linear geometries is in a suitable sense dense in the set of all geometries 
relevant for the path integral (probably a fairly mild assumption), and
2) that we are using a correct measure on the set of geometries. This is 
a more questionable assumption since we do not even know whether such a measure 
exists. Here one has to take a pragmatic attitude in order to make progress. 
We will simply examine the outcome of our construction and try to judge 
whether it is promising.  

Secondly, our scheme is background-independent. No distinguished geometry, 
accompanied by quantum fluctuations, is put in by hand. 
If the CDT-regularized theory is to be taken seriously 
as a potential theory of quantum gravity, there has to be a
region in the space spanned by the bare coupling constants where the geometry 
of spacetime bears some resemblance with the kind of universe we
observe around us. That is, the theory should create dynamically an effective background 
geometry around which there are (small) quantum fluctuations.
This is a very nontrivial property of the theory and one we are going
to investigate in detail in the present piece of work. 
New computer simulations presented here confirm in a much more direct
way the indirect evidence for such a scenario which we provided
earlier in \cite{emerge,semi}. 
They establish the de Sitter nature of the background spacetime, 
quantify the fluctuations around it,
and set a physical scale for the universes we are dealing with.
The main results of our investigation, without the numerical details, 
were announced in \cite{agjl} (see also \cite{desitter}).   

The rest of the article is organized as follows. In Sec.\ \ref{cdt}
we describe briefly the regularization method of quantum gravity named CDT
and the set-up of the computer simulations. In Sec.\ \ref{S4} we 
present the evidence for an effective background geometry corresponding
to the four-dimensional sphere $S^4$, i.e.\ Euclidean de Sitter spacetime. 
Sec.\ \ref{effective} 
deals with the reconstruction of an effective action for the scale factor of the universe
from the computer data, and in Sec.\ \ref{fluctuations}
we analyze the quantum fluctuations around the ``classical'' $S^4$-solution.
Sec.\ \ref{S3} contains an analysis of the geometry of the spatial slices
of our computer-generated universe. In Sec.\ \ref{size} we determine
the physical sizes of our universes expressed in Planck lengths and try
to follow the flow of the gravitational coupling constant of
the effective action under a change of the bare coupling constants of
the bare, ``classical'' action used in the path integral. 
Finally we discuss the results, their interpretation and future perspectives
of the CDT-quantum gravity theory in Sec.\ \ref{discussion}.

\section{Causal Dynamical Triangulations (CDT)}\label{cdt}

The approach of causal dynamical triangulations stands in the tradition of 
\cite{teitelboim}, which advocated that in a gravitational path integral with the 
correct, Lorentzian signature of spacetime one should sum 
over causal geometries only. More specifically, we adopted this idea when it
became clear that 
attempts to formulate a {\it Euclidean} nonperturbative quantum gravity
theory run into trouble in spacetime dimension $d$ larger than two.
At the same time, such a causal reformulation results in a path integral which 
relates more closely to canonical formulations of quantum gravity.

This implies that we start from Lorentzian simplicial spacetimes with $d=4$ and insist
that only causally well-behaved geometries appear in the
(regularized) Lorentzian path integral. 
A crucial property of our explicit construction is that 
each of the configurations allows for a rotation to Euclidean signature.
We rotate to a Euclidean regime in order to perform the sum over geometries
(and rotate back again afterwards if needed).
We stress here that although the sum is performed over geometries 
with Euclidean signature, it is different from what one would 
obtain in a theory of quantum gravity based ab initio on Euclidean spacetimes.
The reason is that not all Euclidean geometries with a given topology are 
included in the ``causal" sum since in general they have no correspondence 
to a causal Lorentzian geometry.
 
How do we construct the class of piecewise linear geometries used 
in the Lorentzian path integral (see \cite{ajl4d} for a detailed
description)? The most important assumption is the existence of 
a global proper-time foliation. We assume that the spacetime
topology is that of $I \times \Sigma^{(3)}$, where $\Sigma^{(3)}$ denotes 
an arbitrary three-dimensional manifold.  
In what follows, we will for simplicity study the case of 
the simplest spatial topology $\Sigma^{(3)}=S^3$, that of a three-sphere.
The compactness of $S^3$ obviates the discussion of spatial boundary
conditions for the universe.
The spatial geometry at each discrete proper-time step $t_n$ is represented
by a triangulation of $S^3$, made up of equilateral spatial tetrahedra with 
squared side-length $\ell_s^2\equiv a^2 >0$. In general, the number $N_3(t_n)$ of tetrahedra and how 
they are glued together to form a piecewise flat three-dimensional manifold
will vary with each time-step $t_n$. In order to obtain a four-dimensional triangulation,
the individual three-dimensional slices must still be connected in a causal way, 
preserving the $S^3$-topology at all intermediate times $t$ between 
$t_n$ and $t_{n+1}$.\footnote{This implies the absence of branching of the spatial
universe into several disconnected pieces, so-called {\it baby universes}, 
which (in Lorentzian signature) would 
inevitably be associated with causality violations in the form of degeneracies in
the light cone structure, as has been discussed elsewhere (see, for example, \cite{causality}).}
This is done by connecting each tetrahedron belonging to the 
triangulation at time $t_n$ to a vertex belonging to the triangulation
at time $t_{n+1}$ by means of a four-simplex which has four time-like links 
of length-squared $\ell_t^2\equ -\alpha \ell_s^2$, $\alpha >0$, 
interpolating between the adjacent slices
(a so-called (4,1)-simplex). In addition, a triangle in the triangulation
at time $t_n$ can be connected to a link in the triangulation at $t_{n+1}$ via
a four-simplex with six time-like links (a so-called (3,2)-simplex), 
again with $\ell_t^2\equ -\alpha \ell_s^2$. 
Conversely, one can connect a link at $t_n$
to a triangle at $t_{n+1}$ to create a (2,3)-simplex 
and a vertex at $t_n$ to a tetrahedron at $t_{n+1}$ to create a (1,4)-simplex.
One can interpolate between subsequent triangulations of $S^3$ at $t_n$ and $t_{n+1}$ 
in many distinct ways compatible with the topology $I\times S^3$ of the four-manifold.
All these possibilities are summed over in the CDT path integral.
The explicit rotation to Euclidean 
signature is done by performing the rotation $\a \to -\a$ in the complex lower
half-plane, $|\a| > 7/12$, such that we have 
$\ell_t^2 = |\a| \ell_s^2$ (see \cite{ajl4d} for a discussion).

The Einstein-Hilbert action $S^{\rm EH}$ has a natural geometric implementation
on piecewise linear geometries in the form of the Regge action. 
This is given by the sum of the so-called deficit 
angles around the two-dimensional ``hinges" (subsimplices in the form of triangles), 
each multiplied with the volume
of the corresponding hinge. In view of the fact that we are dealing with piecewise
linear, and not smooth metrics, there is no unique ``approximation'' to the usual 
Einstein-Hilbert action, and one could in principle work with a different form of the
gravitational action. We will stick with the Regge action, which takes on a very 
simple form in our case, where  
the piecewise linear manifold is constructed from just two different types of building blocks.
After rotation to Euclidean signature one obtains for the action
(see \cite{blp} for details)
\bea
S_E^{\rm EH}&=& \frac{1}{16\pi^2 G} \int d^4x \sqrt{g} (-R+2\La) \nonumber \\
 \to S_E^{\rm Regge}&=& -(\kappa_0+6\Delta) N_0+\kappa_4 (N_{4}^{(4,1)}+N_{4}^{(3,2)})+
\Delta (2 N_{4}^{(4,1)}+N_{4}^{(3,2)}),
\label{actshort}
\eea 
where $N_0$ denotes the total number of vertices in the four-dimensional 
triangulation and $N_{4}^{(4,1)}$ and $N_{4}^{(3,2)}$ denote 
the total number of the four-simplices described above, i.e.\ 
the total number of (4,1)-simplices {\it plus} (1,4)-simplices and
the total number of (3,2)-simplices {\it plus} (2,3)-simplices, respectively, so that the
total number $N_4$ of four-simplices is $N_4=N_{4}^{(4,1)}+N_{4}^{(3,2)}$. 
The dimensionless coupling constants $\k_0$ and $\k_4$ are 
related to the bare gravitational and bare cosmological coupling constants,
with appropriate powers of the lattice spacing $a$ already absorbed into 
$\k_0$ and $\k_4$. The {\it asymmetry parameter}
$\Del$ is related to the parameter $\a$ introduced above, which describes 
the relative scale
between the (squared) lengths of space- and time-like links. 
It is both convenient and natural to keep track of this parameter in our set-up, which from the
outset is not isotropic in time and space directions, see again \cite{blp} 
for a detailed discussion. Since we will in the following work with the path integral
after Wick rotation, let us redefine $\tilde \a:=-\a$ \cite{blp}, which is positive in the Euclidean
domain.\footnote{The most symmetric choice is $\tilde\a=1$, corresponding to vanishing
asymmetry, $\Delta=0$.} For future reference, the Euclidean four-volume of our universe 
for a given choice of $\tilde\a$ is given by 
\beq\label{vol1}
V_4 = C_4\, a^4\;\Big(\frac{\sqrt{8\tilde\a -3}}{\sqrt{5}}\; N_4^{(4,1)} +  
\frac{\sqrt{12\tilde\a -7}}{\sqrt{5}}\; N_4^{(3,2)} \Big),
\eeq
where $C_4 = \sqrt{5}/96$ is the four-volume of an equilateral four-simplex 
with edge length $a=1$ (see \cite{ajl4d} for 
details). It is convenient to rewrite expression \rf{vol1} as 
\beq\label{vol2}
V_4 = \tilde{C}_4(\xi)\, a^4\, N_4^{(4,1)} =  
\tilde{C}_4(\xi)\, a^4 \, N_4/(1+\xi),
\eeq
where $\xi$ is the ratio
\beq\label{vol3}
\xi = N_4^{(3,2)}/N_4^{(4,1)},
\eeq
and $\tilde{C}_4(\xi)\, a^4$ is a measure of the ``effective four-volume''
of an ``average'' four-simplex.
In computing \rf{actshort}, we have assumed that the spacetime manifold
is compact without boundaries, otherwise appropriate
boundary terms must be added to the action. 

The path integral or partition function for the CDT version of quantum gravity is now
\beq\label{2.1}
Z(G,\La) = \int \cD [g] \; \e^{-S_E^{\rm EH}[g]} ~~~\to~~~ 
Z(\k_0,\k_4,\Del) = 
\sum_{\cT} \frac{1}{C_\cT} \; \e^{-S_E(\cT)},
\eeq
where the summation is over all causal triangulations $\cT$ of the kind 
described above, and we have dropped the superscript ``Regge" on the discretized action. 
The factor $1/C_\cT$ is a symmetry factor, given by the order of 
the automorphism group of the triangulation $\cT$. 
The actual set-up for the simulations is as
follows. We choose a fixed number $N$ of spatial slices at proper times $t_1$,
$t_2= t_1 \plu a_t$, up to $t_N \equ t_1 \plu (N\mi 1) a_t$, 
where $\Delta t\equiv a_t$ is the discrete lattice
spacing in temporal direction and $T = N a_t$ the 
total extension of the universe in proper time. For convenience we identify $t_{N+1}$
with $t_1$, in this way imposing the topology $S^1\times S^3$ rather
than $I\times S^3$. This choice does not affect physical results,
as will become clear in due course.

Our next task is to {\it evaluate} the nonperturbative sum in \rf{2.1}, if possible, analytically.
Although this can be done in spacetime dimension $d=2$ (\cite{al}, and see \cite{alwz} for 
recent developments) and at least partially in $d=3$ \cite{worm,blz}, an analytic solution in
four dimensions is currently out of reach. However, we are in the fortunate situation that
$Z(\kappa_0,\kappa_4,\Delta)$ can be studied quantitatively with the help of Monte
Carlo simulations. The type of algorithm needed to update 
the piecewise linear geometries has been around for a while,
starting from the use of dynamical triangulations in bosonic string
theory (two-dimensional Euclidean triangulations) \cite{adf,david,migdal} 
and was later extended to their application in Euclidean four-dimensional quantum 
gravity \cite{aj,migdal1}.
In \cite{ajl4d} the algorithm was modified to accomodate the geometries of the CDT set-up.
Note that the algorithm is such that it takes the symmetry factor $C_\cT$ into account 
automatically. 

We have performed extensive Monte Carlo simulations of the 
partition function $Z$ for a number of  values of the bare coupling constants.
As reported in \cite{blp}, there are regions of the coupling 
constant space which do not appear relevant for continuum physics in
that they seem to suffer from problems similar to the ones found
earlier in {\it Euclidean} quantum gravity constructed in terms of dynamical
triangulations, which essentially led to its abandonement in $d>2$.  
Namely, when the (inverse, bare) gravitational coupling $\k_0$ is sufficiently 
large, the Monte Carlo simulations exhibit
a sequence in time direction of small, disconnected universes, none of 
them showing any sign of the scaling one would 
expect from a macroscopic universe. 
We believe that this phase of the system is a Lorentzian version of the 
branched polymer phase of Euclidean quantum gravity. 
By contrast, when $\Del$ is sufficiently small the simulations
reveal a universe with a vanishing temporal extension of only a few 
lattice spacings, ending both in past and future in a vertex of very high order, 
connected to a large fraction of all vertices. This phase is most likely related
to the so-called crumpled phase of Euclidean quantum gravity.
The crucial and new feature of the quantum superposition in terms of 
{\it causal} dynamical triangulations is the appearance of a region 
in coupling constant space which is different and interesting and 
where continuum physics may emerge. 
It is in this region that we have performed the simulations reported in this article,
and where previous work has already uncovered a number of intriguing
physical results \cite{emerge,semi,blp,spectral}. 

In the Euclideanized setting the value of the cosmological constant 
determines the spacetime volume $V_4$ since the two appear in the action as
conjugate variables. We therefore have $\la V_4\ra \sim G/\La$ in a continuum 
notation, where 
$G$ is the gravitational coupling constant and $\La$ the cosmological 
constant. In the computer simulations it is more 
convenient to keep the four-volume fixed or partially fixed. We will
implement this by fixing the total number of 
four-simplices of type $N_4^{(4,1)}$ or, equivalently, the total number $N_3$
of tetrahedra making up the spatial $S^3$ triangulations at
times $t_i$, $i= 1,\dots,N$,
\beq\label{2.3}
N_3 =\sum_{i=1}^N N_3(t_i) = \oh N_4^{(4,1)}.
\eeq 
We know from the simulations
that in the phase of interest $\la N_4^{(4,1)} \ra \propto \la N_4^{(3,2)} \ra$ as the 
total volume is varied \cite{blp}. This
effectively implies that we only have two bare coupling constants 
$\k_0,\Del$ in \rf{2.1}, while we compensate by hand for the coupling
constant $\k_4$ by studying the partition function $Z(\k_0,\Del;N_4^{(4,1)})$
for various $N_4^{(4,1)}$. 
To keep track of the ratio $\xi(\k_0,\Del)$ between the expectation value 
$\la N_4^{(3,2)} \ra$ and $N_4^{(4,1)}$, which depends weakly on the coupling constants, 
we write (c.f. eq.\ \rf{vol3})
\beq\label{ratio}
 \la N_4 \ra = N_4^{(4,1)} + \la N_4^{(3,2)}\ra = 
N_4^{(4,1)}(1+\xi(\k_0,\Del)).
\eeq
For all practical purposes we can regard $N_4$ in
a Monte Carlo simulation as fixed. The relation between the partition function 
we use and the partition function with variable four-volume is 
given by the Laplace transformation
\beq\label{2.3x}
Z(\k_0,\k_4,\Del) = \int_0^\infty \d N_4 \; \e^{-\k_4 N_4} \; 
Z(\k_0,N_4,\Del), 
\eeq
where strictly speaking the integration over $N_4$ should be replaced by a summation
over the discrete values $N_4$ can take.

\section{The macroscopic de Sitter universe\label{S4}}  

The Monte Carlo simulations referred to above will generate a sequence of 
spacetime histories. An individual spacetime history is not an
observable, in the same way as a path $x(t)$ of a particle in the 
quantum-mechanical path integral is not. However, 
it is perfectly legitimate to talk about the {\it expectation value} $\la x(t) \ra$ as 
well as the {\it fluctuations around} $\la x(t) \ra$. Both of these quantities
are in principle calculable in quantum mechanics. 

Obviously, there are many more dynamical variables in quantum gravity than 
there are in the particle case. We can still imitate the quantum-mechanical situation 
by picking out a particular one, for example, 
the spatial three-volume $V_3(t)$ at proper time $t$. We 
can measure both its expectation value $\la V_3(t)\ra $ as well as fluctuations
around it. The former gives us information about the large-scale  
``shape'' of the universe we have created in the computer. In this
section, we will describe the measurements of $\la V_3(t)\ra $, keeping a more
detailed discussion of the fluctuations to Sec.\ \ref{fluctuations} below.

A ``measurement'' of $V_3(t)$ consists of a table $N_3(i)$, where
$i=1,\ldots,N$ denotes the number of time-slices. Recall from Sec.\ \ref{cdt}
that the sum over slices $\sum_{i=1}^N N_3(i)$ is kept constant. The 
time axis has a total length of $N$ time steps, where $N=80$ in the 
actual simulations, and we have cyclically identified time-slice $N+1$ with
time-slice 1. 

What we observe in the simulations is that for the range of discrete volumes
$N_4$ under study the universe does {\it not} extend
(i.e. has appreciable three-volume) over the entire time axis, but rather is
localized in a region much shorter than 80 time slices. 
Outside this region the spatial extension $N_3(i)$ will be minimal, consisting of the 
minimal number (five) of tetrahedra needed to 
form a three-sphere $S^3$, plus occasionally a few more 
tetrahedra.\footnote{This
kinematical constraint ensures that the triangulation remains a {\it simplicial
manifold} in which, for example, two $d$-simplices are not allowed to have more than 
one $(d-1)$-simplex in common.}
This thin ``stalk" therefore
carries little four-volume and in a given simulation we can for most practical purposes
consider the total four-volume of the remainder, the extended universe, as fixed. 

In order to perform a meaningful average over geometries which explicitly refers to 
the extended part of the universe, 
we have to remove the translational zero mode which is present. 
During the Monte Carlo 
simulations the extended universe will fluctuate in shape and its
{\it centre of mass} (or, more to the point, its {\it centre of volume})
will perform a slow random walk along the time axis. 
Since we are dealing with a circle (the compactified time axis), the centre of volume is not 
uniquely defined (it is clearly arbitrary for a constant volume distribution), and we must 
first define what we mean by such a concept. 
Here we take advantage of the empirical fact that our dynamically generated
universes decompose into an extended piece and a stalk, 
with the latter 
containing less than one per cent of the total volume. We are clearly interested
in a definition such that the centre of volume of a given configuration lies in the
centre of the extended region. One also expects that any sensible definition
will be unique up to contributions related 
to the stalk and to the discreteness of the time steps. In total
this amounts to an ambiguity of the centre of volume of one lattice step in the time
direction.

In analyzing the computer data we have chosen one specific definition
which is in accordance with the discussion above\footnote{Explicitly, 
we consider the quantity
$$
CV(i') = \left| \sum_{i=-N/2}^{N/2-1} (i+0.5) N_3(1 + {\rm mod}(i' + i - 1, N) ) \right|
$$
and find the value of $i' \in \{1,\dots,N \}$ for which $CV(i')$ is smallest. 
We denote this $i'$ by $i_{cv}$. If there is
more than one minimum, we choose the value which has the largest 
three-volume $N_3(i')$. Let us stress that
this is just one of many definitions of $i_{cv}$. All other sensible 
definitions will for the type of configurations considered here
agree to within one lattice spacing.}.
Maybe surprisingly, it turns out that the inherent ambiguity in the choice of a definition
of the centre of volume -- even if it is only of the order of one lattice
spacing -- will play a role later on in our analysis of the quantum fluctuations. 
For each universe used in the 
measurements (a ``path" in the gravitational path integral) 
we will denote the centre-of-volume time coordinate calculated
by our algorithm by $i_{cv}$. From now on,
when comparing different universes, i.e.\ when performing ensemble
averages, we will redefine the temporal coordinates according to
\beq\label{cm}
N_3^{new} (i) =  N_3(1 + {\rm mod}(i + i_{cv} - 1, N)),
\eeq 
such that the centre of volume is located at 0. 

\begin{figure}
\centerline{\scalebox{0.75}{\rotatebox{0}{\includegraphics{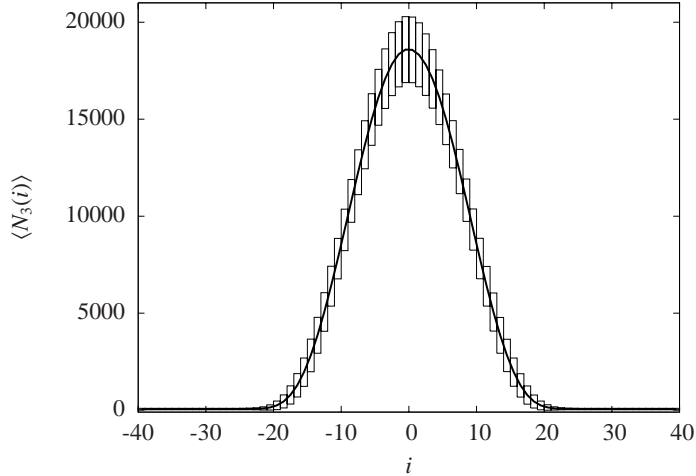}}}}
\caption{\label{fig1} Background geometry $\langle N_3(i)\rangle$: 
MC measurements for fixed $N_4^{(4,1)}= 160.000$ ($N_4=362.000$)
and best fit \rf{n1} yield indistinguishable curves at given plot resolution. 
The bars indicate the average size of quantum fluctuations.}
\end{figure}

Having defined in this manner the centre of volume along the 
time-direction of our spacetime configurations we can now 
perform superpositions of such configurations and 
define the average $\la N_3(i)\ra $ as a function of the discrete time $i$.
The results of measuring the average discrete 
spatial size of the universe at various 
discrete times $i$ are illustrated 
in Fig.\ \ref{fig1} and can be succinctly summarized by the formula
\beq\label{n1}
N_3^{cl}(i):= \la N_3(i)\ra  = \frac{N_4}{2(1+\xi)}\;\frac{3}{4} \frac{1}{s_0 N_4^{1/4}}  
\cos^3 \left(\frac{i}{s_0 N_4^{1/4}}\right),~~~s_0\approx 0.59,
\eeq
where $N_3(i)$ denotes the number of three-simplices in the spatial slice 
at discretized time $i$ and $N_4$ the 
total number of four-simplices in the entire universe. Since we are 
keeping $N_4^{(4,1)}$ fixed in the simulations and since $\xi$ changes
with the choice of bare coupling constants, it is sometimes 
convenient to rewrite \rf{n1} as
\beq\label{n1x}
N_3^{cl}(i) = \oh N_4^{(4,1)}\;\frac{3}{4} 
\frac{1}{\tilde{s}_0 (N_4^{(4,1)})^{1/4}}  
\cos^3 \left(\frac{i}{\tilde{s}_0 (N_4^{(4,1)})^{1/4}}\right),
\eeq
where $\tilde{s}_0$ is defined by
$\tilde{s}_0 (N_4^{(4,1)})^{1/4} = s_0 N_4^{1/4}$.
Of course, formula \rf{n1} is only valid in the extended part of the universe 
where the spatial three-volumes are larger than the minimal cut-off size.

The data shown in Fig.\ \ref{fig1} have been collected at the particular values  
$(\k_0,\Del) = (2.2,0.6)$ of the bare coupling constants and for 
$N_4= 362.000$ (corresponding to $N_4^{(4,1)} = 160.000$). 
For these values of $(\k_0,\Del)$
we have verified relation \rf{n1} for $N_4$ ranging from 45.500 to 362.000
building blocks (45.500, 91.000, 181.000 and 362.000). 
After rescaling the time and volume variables by suitable powers of $N_4$ 
according to relation (\ref{n1}), and plotting them in the same way as 
in Fig.\ \ref{fig1}, one finds almost total
agreement between the curves for different spacetime volumes.\footnote{By
contrast, the quantum fluctuations indicated in 
Fig.\ \ref{fig1} as vertical bars {\it are} volume-dependent and will be the larger the
smaller the total four-volume, see Sec.\ \ref{fluctuations} below for details.}
Eq.\ \rf{n1} shows that
spatial volumes scale according to $N_4^{3/4}$ and time intervals
according to $N_4^{1/4}$, as one would expect for
a genuinely {\it four}-dimensional spacetime. This strongly suggests
a translation of \rf{n1} to a continuum notation.
The most natural identification is given by  
\beq\label{n2}
\sqrt{g_{tt}}\; V_3^{cl}(t) = V_4 \;
\frac{3}{4 B} \cos^3 \left(\frac{t}{B} \right),
\eeq
where we have made the identifications
\beq\label{n3}
\frac{t_i}{B} = \frac{i}{s_0 N_4^{1/4}}, ~~~~
\Del t_i \sqrt{g_{tt}}\;V_3(t_i) = 2 \tC_4 N_3(i) a^4,
\eeq
such that we have
\beq\label{n3z}
\int dt \sqrt{g_{tt}} \; V_3(t) = V_4.
\eeq
In \rf{n3}, $\sqrt{g_{tt}}$ is the constant proportionality 
factor between the time
$t$ and genuine continuum proper time $\tau$, $\tau=\sqrt{g_{tt}}\; t$.
(The combination $\Del t_i\sqrt{g_{tt}}V_3$ contains $\tC_4$, related to the 
four-volume of a four-simplex rather than the three-volume corresponding
to a tetrahedron, because its time integral must equal $V_4$).
Writing $V_4 =8\pi^2 R^4/{3}$, and $\sqrt{g_{tt}}=R/B$,
eq.\ \rf{n2} is seen to describe 
a Euclidean {\it de Sitter universe} 
(a four-sphere, the maximally symmetric space for 
positive cosmological constant)
as our searched-for, dynamically generated background geometry!
In the parametrization of \rf{n2} 
this is the classical solution to the action
\beq\label{n5}
S= \frac{1}{24\pi G} \int d t \sqrt{g_{tt}}
\left( \frac{ g^{tt}\dot{V_3}^2(t)}{V_3(t)}+k_2 V_3^{1/3}(t)
-\l V_3(t)\right),
\eeq
where $k_2= 9(2\pi^2)^{2/3}$ and  $\l$ is a Lagrange multiplier,
fixed by requiring that the total four-volume 
be $V_4$, $\int d t \sqrt{g_{tt}} \;V_3(t) = V_4$. 
Up to an overall sign, this is precisely 
the Einstein-Hilbert action for the scale 
factor $a(t)$ of a homogeneous, isotropic universe
(rewritten in terms of the spatial three-volume $V_3(t) =2\pi^2 a(t)^3$), 
although we of course never put any such
simplifying symmetry assumptions into the CDT model. 

For a fixed, finite four-volume $V_4$ and when applying scaling arguments
it can be convenient to rewrite \rf{n5} in terms of dimensionless units by
introducing $s=t/V_4^{1/4}$ and $V_3(t) = V_4^{3/4} v_3(s)$, in which case
\rf{n5} becomes
\beq\label{n5a}
S =  \frac{1}{24\pi}\; \frac{\sqrt{V_4}}{G} \int d s \sqrt{g_{ss}}
\left( \frac{ g^{ss}\dot{v_3}^2(s)}{v_3(s)}+k_2 v_3^{1/3}(s) \right),
\eeq
now assuming that $\int ds \sqrt{g_{ss}}\; v_3(s) = 1$, and with $g_{ss}\equiv g_{tt}$.
A discretized, dimensionless version of \rf{n5} is
\beq\label{n7b}
S_{discr} =
k_1 \sum_i \left(\frac{(N_3(i+1)-N_3(i))^2}{N_3(i)}+
\tilde{k}_2 N_3^{1/3}(i)\right),
\eeq
where $\tilde{k}_2\propto k_2$.
This can be seen by applying the scaling \rf{n1}, namely, $N_3(i) = N_4^{3/4} n_3(s_i)$
and $s_i = i/N_4^{1/4}$. With this scaling, the action \rf{n7b} becomes
\beq\label{n7bb}
 S_{discr} =
k_1 \sqrt{N_4} \sum_i  \Delta s  \left(\frac{1}{n_3(s_i)}     
\left(\frac{n_3(s_{i+1})-n_3(s_i)}{\Delta s}\right)^2+
\tilde{k}_2 n_3^{1/3}(s_i)\right),
\eeq
where $\Delta s = 1/N^{1/4}$, and therefore has the same form as \rf{n5a}.
This enables us to finally conclude that the identifications \rf{n3} when used in 
the action \rf{n7b} lead 
na\"ively to the continuum expression \rf{n5} under the identification
\beq\label{n7c}
G = \frac{a^2}{k_1} \frac{\sqrt{\tC_4}\; \ts_0^2}{3\sqrt{6} }.
\eeq
\begin{figure}[th]
\centerline{\scalebox{1.00}{\rotatebox{0}{\includegraphics{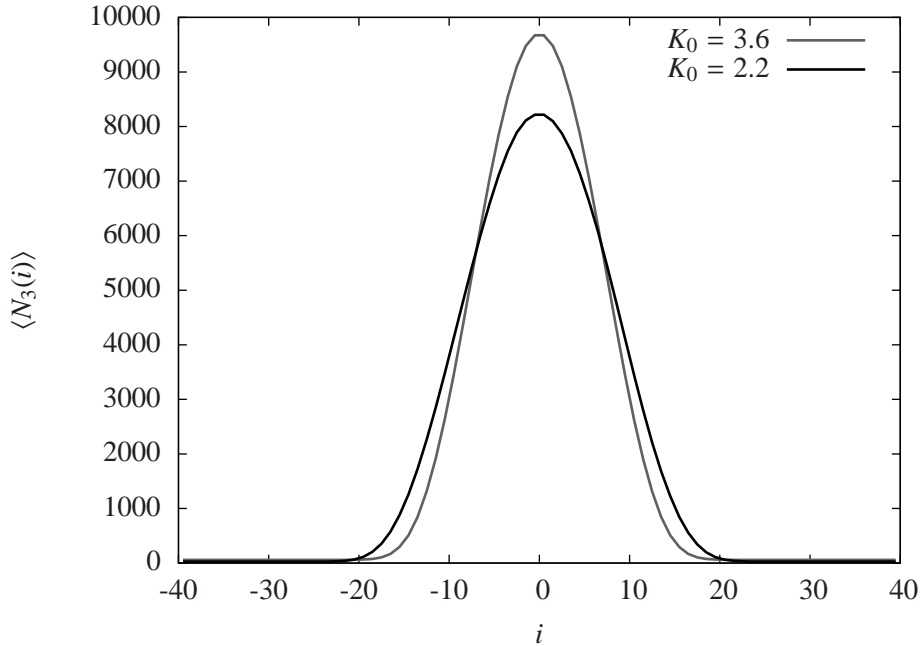}}}}
\caption{\label{fig2} The measured average shape $\la N_3(i)\ra $ 
of the quantum universe at $\Del = 0.6$, for $\k_0$
= 2.2 (broader distribution) and $\k_0$ = 3.6 (narrower distribution), taken
at $N_4^{(4,1)}=160.000.$}
\end{figure}

Next, let us comment on the universality of these results. 
First, we have checked that they are not
dependent on the particular definition of time-slicing we have 
been using, in the following sense. By construction of the
piecewise linear CDT-geometries we have at each integer time step $t_i= i\, a_t $
a spatial surface consisting of $N_3(i)$ tetrahedra. Alternatively, one
can choose as reference slices for the measurements of the spatial volume
non-integer values of time, for example, all time slices at discrete times $i-1/2$,
$i=1,2,...$ . In this case the ``triangulation" of the spatial three-spheres consists of
tetrahedra -- from cutting a (4,1)- or a (1,4)-simplex half-way -- and ``boxes",
obtained by cutting a (2,3)- or (3,2)-simplex (the geometry of this is worked out
in \cite{dl}). We again find a relation
like \rf{n1} if we use the total number of spatial building blocks
in the intermediate slices (tetrahedra+boxes) instead of just the 
tetrahedra. 

\begin{figure}[th]
\centerline{\scalebox{1.00}{\rotatebox{0}{\includegraphics{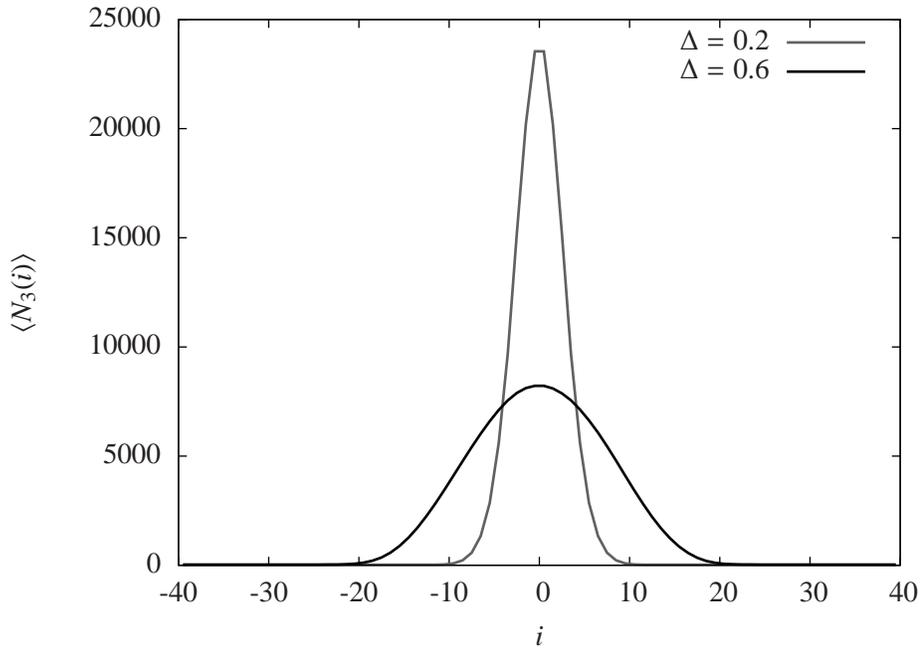}}}}
\caption{\label{fig3} The measured average shape $\la N_3(i)\ra $ of the quantum
universe at $\k_0 = 2.2$, for $\Del$
= 0.6 (broad distribution) and $\Del$= 0.2 (narrow distribution), both taken at
$N_4^{(4,1)}=160.000$.}
\end{figure}

Second, we have repeated the measurements for other values of the bare 
coupling constants. As long as we stay in the phase where 
an extended universe is observed (called ``phase C" in ref.\ \cite{blp}), 
a relation like \rf{n1} remains
valid. In addition, the value of $s_0$, defined in eq.\ \rf{n1}, is almost unchanged 
until we get close to the phase transition
lines beyond which the extended universe disappears.
Fig.\ \ref{fig2} shows the average shape $\la N_3(t)\ra $ for $\Del = 0.6$ and for $\k_0$
equal to 2.2 and 3.6. Only for the values of $\k_0$ around 3.6 and larger will the
measured $\la N_3(t)\ra $ differ significantly from the value at 2.2.
For values larger than 3.8 (at $\Del = 0.6$), the
universe will disintegrate into a number of small and disconnected components 
distributed randomly along the time axis, and one can no longer fit the distribution 
$\la N_3(t)\ra $ to the formula \rf{n1}. 
Fig.\ \ref{fig3} shows the average shape $\la N_3(t)\ra $ for $\k_0= 2.2$ and
$\Del $ equal to 0.2 and 0.6. Here the value $\Del = 0.2$ is close to the
phase transition where the extended universe will flatten out to a 
universe with a time extension of a few lattice spacings only.
Later we will show that while $s_0$ is almost unchanged, the constant 
$k_1$ in \rf{n7b}, which governs the quantum fluctuations around the 
mean value $\la N_3(t)\ra $, is more sensitive to a change
of the bare coupling constants, in particular in the case where we
change $\k_0$ (while leaving $\Del$ fixed).

\section{Constructive evidence for the effective action\label{effective}}

While the functional form \rf{n1} for the three-volume fits the data perfectly and
the corresponding continuum effective action \rf{n5} reproduces
the continuum version \rf{n2} of \rf{n1}, it is still of interest to 
check to what extent one can reconstruct the discretized
version \rf{n7b} of the continuum action \rf{n5} from the data explicitly. Stated differently, 
we would like to understand whether there are 
other effective actions which reproduce the data equally well.
As we will demonstrate by explicit construction in this section, there is good evidence 
for the uniqueness of the action \rf{n7b}.

The data we have are two-fold: the measurement of $N_3(i)$, that is, the three-volume
at the discrete time step $i$, and the measurement of 
the three-volume correlator $N_3(i) N_3(j)$. 
Having created $K$ statistically independent configurations $N_3^{(k)} (i)$ by Monte Carlo 
simulation allows us to construct the average
\beq\label{5.1}
\bar{N}_3(i) := \la N_3(i) \ra \cong \frac{1}{K} \sum_k N_3^{(k)} (i),
\eeq
where the superscript in $(\cdot)^{(k)}$ denotes the result 
of the k'th configuration sampled, 
as well as the covariance matrix
\beq\label{5.2}
C(i,j) \cong \frac{1}{K} \sum_k 
(N^{(k)}_3 (i) -\bar{N}_3(i))(N^{(k)}_3 (j) -\bar{N}_3(j)).
\eeq   
Since we have fixed the sum $\sum_{i=1}^{N} N_3(i)$ (recall that $N$ denotes the
fixed number of time steps in a given simulation), the covariance
matrix has a zero mode, namely, the constant vector $e^{(0)}_i$,
\beq\label{5.3}
\sum_i C(i,j)e^{(0)}_j = 0,~~~~e^{(0)}_i = 1/\sqrt{N}\quad \forall i.
\eeq
A spectral decomposition of the symmetric covariance matrix gives
\beq\label{5.4}
\hat{C} = \sum_{a=1}^{N-1} \l_a | e^{(a)}\ra \la e^{(a)}|,
\eeq
where we assume the $N-1$ other eigenvalues of the covariance matrix 
$\hat{C}_{ij}$ are different from zero. We now define the ``propagator" 
$\hat{P}$ as
the inverse of $\hat{C}$ on the subspace orthogonal to the zero mode 
$e^{(0)}$ , that is,
\beq\label{5.5}
\hat{P} = \sum_{a=1}^{N-1} \frac{1}{\l_a} | e^{(a)}\ra \la e^{(a)}| =
(\hat{C} +\hat{A})^{-1}-\hat{A},~~~\hat{A}=| e^{(0)}\ra \la e^{(0)}|.
\eeq

We now assume we have a discretized action which can 
be expanded around the expectation value $\bar{N}_3(i)$
according to
\beq\label{5.6}
 S_{discr}[\bar{N}+n] = S_{discr}[\bar{N}] +
\oh \sum_{i,j} n_i \hat{P}_{ij} n_j  +O(n^3).
 \eeq
If the quadratic approximation describes 
the quantum fluctuations around the expectation value $\bar{N}$ well,
the inverse of $\hat{P}$ will be a good 
approximation to the covariance matrix.
Conversely, still assuming the quadratic approximation gives 
a good description of the 
fluctuations, the $\hat{P}$ constructed from the covariance matrix will to a 
good approximation allow us to reconstruct the action via \rf{5.6}.

Simply by looking at the inverse $\hat P$ of the measured covariance matrix, 
defined as described
above, we observe that it is to a very good approximation 
small and constant except on the 
diagonal and the entries neighbouring the diagonal. 
We can then decompose it into a ``kinetic'' and a ``potential'' term.
The kinetic part $\hPk$ is defined as the matrix with non-zero 
elements on the diagonal and in the 
neighbouring entries, such that the sum of the elements 
in a row or a column is always zero,
\beq\label{5.6a}
\hPk = \sum_{i=1}^N p_i \hX^{(i)},
\eeq
where the matrix $\hX^{(i)}$ is given by
\beq\label{5.7}
\hX^{(i)}_{jk}= \del_{ij}\del_{ik}+\del_{(i+1)j}\del_{(i+1)k}-\del_{(i+1)j}\del_{ik}-
\del_{ij}\del_{(i+1)k}.
\eeq
Note that the range of $\hPk$ lies by definition in the 
subspace orthogonal to the zero mode.
Similarly, we define the potential term 
as the projection of a diagonal matrix $\hat D$ on the subspace
orthogonal to the zero mode 
\beq\label{5.8}
\hPp = (\hI -\hA) \hD (\hI -\hA) = \sum_{i=1}^{N} u_i \hY^{(i)}.
\eeq
The diagonal matrix $\hD$ and the matrices $\hY^{(i)}$ are defined by
\beq\label{5.9}
\hD_{jk} = u_j \del_{jk},~~~\hY^{(i)}_{jk} = \del_{ij}\del_{ik} -\frac{\del_{ij}+\del_{ik}}{N}+\frac{1}{N^2},
\eeq
and $\hI$ denotes the $N\times N$ unit matrix. 

The matrix $\hP$ is obtained from the numerical data by inverting 
the covariance matrix $\hC$ after subtracting the zero mode, 
as described above. 
We can now try to find the best values of 
the $p_i$'s and $u_i$'s by a least-$\chi^2$ fit\footnote{A $\chi^2$-fit 
of the form \rf{5.10} gives the same weight to each
three-volume $N_3(i)$. One might argue that more weight
should be given to the larger $N_3(i)$ in a configuration
since we are interested in the continuum physics and
not in what happens in the stalk where 
$N_3(i)$ is very small. We have tried various $\chi^2$-fits
with reasonable weights associated with 
the three-volumes $N_3(i)$. The kinetic term, which is the 
dominant term, is insensitive to any (reasonable) weight 
associated with $N_3(i)$. The potential term, which will be analyzed 
below, is more sensitive to the choice of the weight. However,
the general power law dependence reported below is again unaffected
by this choice.} to 
\beq\label{5.10}
\tr \left(\hP - (\hPk+\hPp)\right)^2.
\eeq

\begin{figure}[ht]
\centerline{\scalebox{1.00}{\rotatebox{0}{\includegraphics{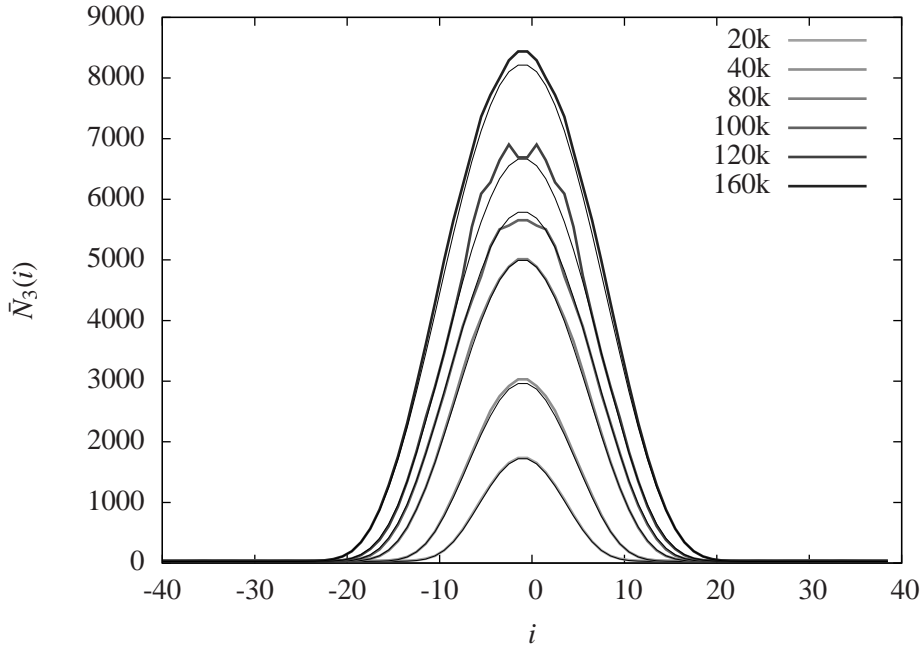}}}}
\caption{\label{fig4} The directly measured expectation values
$\bar{N}_3(i)$ (thick gray curves), compared to the averages $\bar{N}_3(i)$ 
reconstructed from the measured covariance matrix $\hat C$ (thin black curves),
for $\k_0=2.2 $ and $\Del
= 0.6$, at various fixed volumes $N_4^{(4,1)}$. The two-fold symmetry of the interpolated
curves around the central symmetry axis results from an explicit symmetrization of the
collected data.}
\end{figure}

Let us look at the discretized minisuperspace action 
\rf{n7b} which obviously has served as an 
inspiration for the definitions of $\hPk$ and $\hPp$. 
Expanding $N_3(i)$ to second order
around $\bN_3(i)$ one obtains the identifications
\beq\label{5.11}
\bN_3(i) = \frac{2 k_1}{p_i},~~~~~~U''(\bN_3(i)) = -u_i,
\eeq
where $U(N_3(i))=k_1\tilde k_2 N_3^{1/3}(i)$ denotes the potential term in \rf{n7b}. 
We use the fitted coefficients $p_i$
to reconstruct $\bN_3(i)$ and then compare these reconstructed 
values with the averages $\bN_3(i)$ measured directly. 
Similarly, we can use the measured $u_i$'s to 
reconstruct the second derivatives $U''(\bN_3(i))$ and 
compare them to the form $\bN^{-5/3}_3(i)$ coming from \rf{n7b}.

The reconstruction of $\bN_3(i)$ is illustrated in Fig.\ \ref{fig4} for a variety 
of four-volumes $N_4$ and compared with the directly measured
expectation values $\bN_3(i)$. It is seen that the reconstruction works very well and, 
{\it most importantly}, 
the coupling constant $k_1$, which in this 
way is determined independently for each 
four-volume $N_4$ really {\it is} independent of 
$N_4$ in the range of $N_4$'s considered, as should be. 

\begin{figure}[ht]
\centerline{\scalebox{1.00}{\rotatebox{0}{\includegraphics{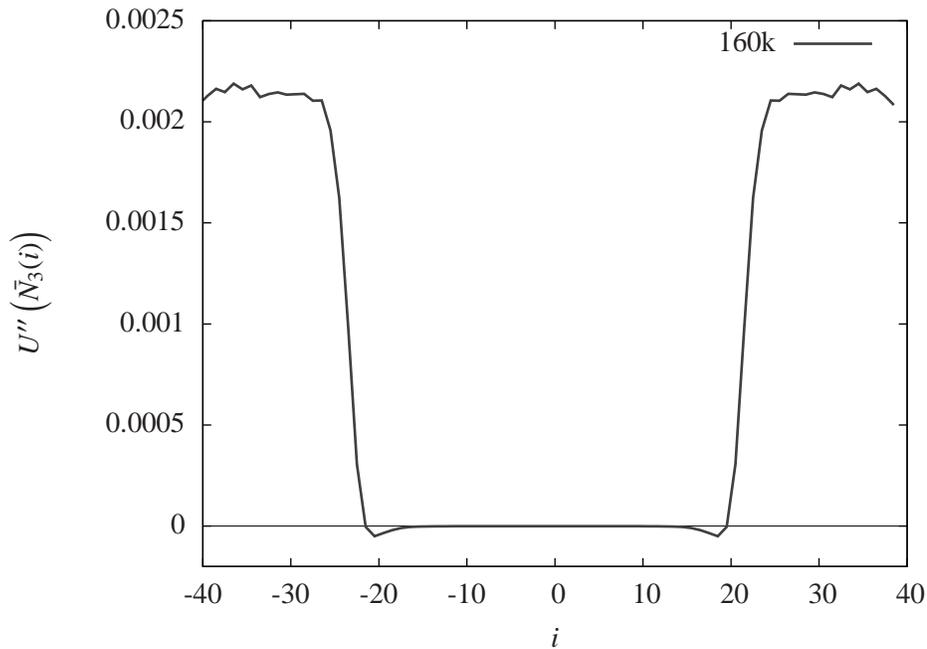}}}}
\caption{\label{fig4a} Reconstructing the second derivative $U''(\bar{N}_3(i))$ from
the coefficients $u_i$, for $\k_0=2.2$
and $\Del= 0.6$ and $N_4^{(4,1)}=160.000$.}
\end{figure}

We will now try to extract the potential $U''(\bN_3(i))$ from the 
information contained in the matrix $\hPp$. The determination of 
$U''(\bN_3(i))$ is not an
easy task as can be understood from Fig.\ \ref{fig4a}, 
which shows the measured coefficients $u_i$ extracted
from the matrix $\hPp$, and which we consider
somewhat remarkable. The interpolated curve makes an
abrupt jump by two orders of magnitude going from the 
extended part of the universe (stretching over roughly 40 time steps) 
to the stalk. The occurrence of this jump is entirely dynamical, no distinction 
has ever been made by hand between stalk and bulk. 

\begin{figure}[ht]
\centerline{\scalebox{1.00}{\rotatebox{0}{\includegraphics{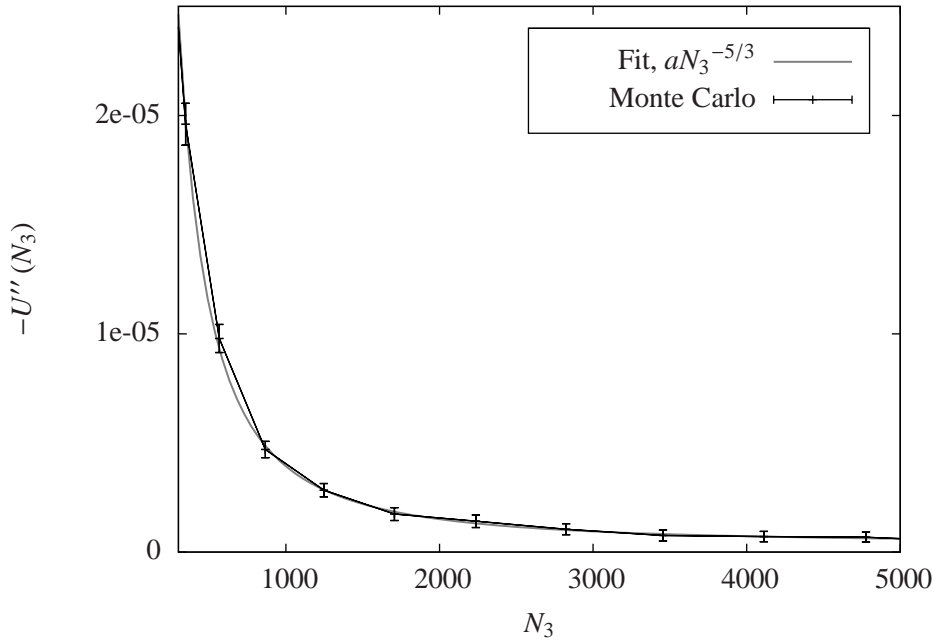}}}}
\caption{\label{fig5} The second derivative 
$-U''(N_3)$ as measured for $N_4^{(4,1)}= 160.000$
and $\k_0 = 2.2$ and $\Del = 0.6$.}
\end{figure}

There are at least two reasons for why it is difficult 
to determine the potential numerically. Firstly, the results 
are ``contaminated" by the presence of the stalk. 
Since it is of cut-off size, its dynamics is dominated
by fluctuations which likewise are of cut-off size. 
They will take the form of short-time sub-dominant contributions
in the correlator matrix $\hC$.
Unfortunately, when we invert $\hC$ to obtain the propagator $\hP$, the
same excitations will correspond 
to the largest eigenvalues and give a very large contribution. 
Although the stalk contribution in the matrix $\hC$ is located 
away from the bulk-diagonal, it can be seen from the appearance of
the $1/N^2$-term in
eqs.\ \rf{5.8} and \rf{5.9} that after 
the projection orthogonal to the 
zero mode the contributions from the stalk will also 
affect the remainder of the geometry in the form of fluctuations
around a small constant value. In deriving Fig.\ \ref{fig5} we have 
subtracted this constant value as best possible. 
However, the {\it fluctuations} of the stalk cannot be subtracted and 
only good statistics can eventually eliminate their effect on the 
behaviour of the extended part of the quantum universe. 
The second (and less serious) reason is that from a 
numerical point of view the potential term is always sub-dominant 
to the kinetic term for the individual spacetime histories in 
the path integral. For instance, consider the simple example of the 
harmonic oscillator. Its discretized action reads
\beq\label{k1}
S = \sum_{i=1}^N  \Del t 
\left[\Big(\frac{x_{i+1}-x_i}{\Del t}\Big)^2 + \om^2 x_i^2\right],
\eeq
from which we deduce that the ratio between the kinetic and potential 
terms will be of order $1/\Del t$ as $\Del t$ tends to zero. This reflects the well-known fact 
that the kinetic term will dominate and 
go to infinity in the limit as $\Del t \to 0$, with
a typical path being nowhere differentiable. 
The same will be true when 
dealing with a more general action like \rf{n5} and its discretized version
\rf{n7b}, where $\Del t$ scales like $\Del t \sim 1/N_4^{1/4}$. Of course,
a {\it classical} solution
will behave differently: there the kinetic term will be 
comparable to the potential term.
However, when extracting the potential term directly from the data, as we 
are doing, one is confronted with this issue. 

The range of the discrete three-volumes $N_3(i)$ in the extended universe is from several 
thousand down to five, the kinematically allowed minimum. However, the behaviour 
for the very small values of $N_3(i)$ near the edge of the extended universe
is likely to be mixed in with 
discretization effects. In order to test whether one really has a $N_3^{1/3}(i)$-term
in the action one should therefore only use values of $N_3(i)$ somewhat larger
than five. This has been done in
Fig.\ \ref{fig5}, where we have 
converted the coefficients $u_i$ from functions of the discrete time steps $i$ 
into functions of the background spatial three-volume $\bN_3(i)$ 
using the identification in \rf{5.11} (the conversion factor can be read off the relevant curve
in Fig.\ \ref{fig4}). 
It should be emphasized that Fig.\ \ref{fig5} is based 
on data from the extended part of the spacetime only; the variation comes entirely
from the central region between times -20 and 20 in Fig.\ \ref{fig4a}, 
which explains why it has been numerically demanding to extract a good signal. 
The data presented in Fig.\ \ref{fig5} were taken at a discrete volume
$N_4^{(4,1)}= 160.000$, and fit well the form $N_3^{-5/3}$, 
corresponding to a potential $\tk_2 N_3^{1/3}$. There is a very small 
residual constant term present in this fit, which presumably is due to the
projection onto the space orthogonal to the zero mode, as already discussed earlier.
In view of the fact that its value is quite close to the noise level with our present statistics,
we have simply chosen to ignore it in the remaining discussion.

\begin{figure}[t]
\centerline{\scalebox{1.00}{\rotatebox{0}{\includegraphics{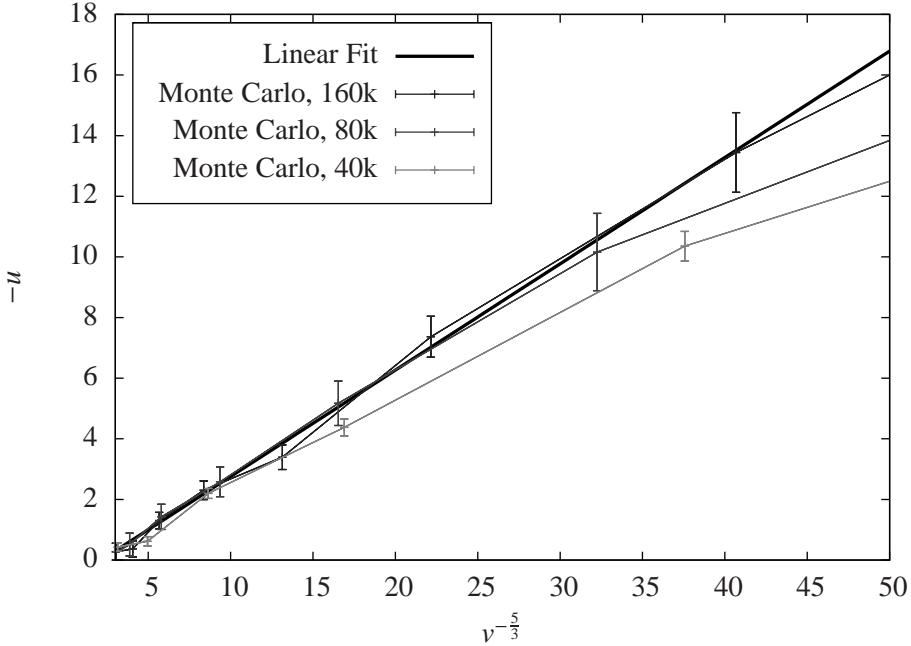}}}}
\caption{\label{fig6} The dimensionless 
second derivative $u= N_4^{5/4} U''(N_3)$ 
plotted against $\n^{-5/3}$, where $\n = N_3/N_4^{3/4}$ is the 
dimensionless spatial volume, for $N_4^{(4,1)}= 40.000$, 
80.000 and 160.000, $\k_0 = 2.2$ and $\Del = 0.6$. One expects a universal
straight line near the origin (i.e. for large volumes) if 
the power law $U(N_3) \propto N^{1/3}$ is correct.}
\end{figure}

Apart from obtaining the correct power $ N_3^{-5/3}$ for the potential
for a given spacetime volume $N_4$, it is equally important that the coefficient 
in front of this term be independent of $N_4$. 
This seems to be the case as is shown in Fig.\ \ref{fig6}, where
we have plotted the measured potentials in terms of reduced, dimensionless
variables which make the comparison between measurements for 
different $N_4$'s easier. --
In summary, we conclude that the data allow us to reconstruct the action
\rf{n7b} with good precision.

Let us emphasize a remarkable aspect of this result.
Our starting point was the Regge action for CDT, as described in Sec.\ \ref{cdt} above.
However, the effective action we have generated dynamically 
by performing the nonperturbative
sum over histories is only indirectly related
to this ``bare'' action. Likewise, the coupling constant $k_1$ which
appears in front of the effective action, and which we view as related to the 
gravitational coupling constant $G$ by eq.\ \rf{n7c} has no obvious direct 
relation to the ``bare'' coupling $\k_0$ appearing in the Regge action \rf{actshort}
and in \rf{2.1}. Nevertheless the leading terms in the 
effective action for the scale factor are precisely the ones presented in \rf{n7b}. 
That a kinetic term with a second-order derivative appears as a 
leading term in an effective action is maybe less surprising, but it is 
remarkable and very encouraging for the entire CDT-quantization program that the 
kinetic term appears in precisely the correct combination with the factor
$N_3(i)^{1/3}$ needed to identify the leading
terms with the corresponding terms in the Einstein-Hilbert action. 
In other words, only 
if these terms are present can we claim to have an effective 
field theory which has anything to do with the standard diffeomorphism-invariant
gravitational theory in the continuum. 
This is neither automatic nor obvious, since our starting point involved both a discretization
and an explicit asymmetry between space and time, and since
the nonperturbative interplay of the local geometric
excitations we are summing over in the path integral is beyond our analytic control.
Nevertheless, what we have found is that at least the leading 
terms in the effective action we have derived dynamically admit an interpretation as the 
standard Einstein term, thus passing a highly nontrivial consistency test.

\section{Fluctuations around de Sitter space\label{fluctuations}}

We have shown that the action \rf{n7b} 
gives a very good description of the measured shape
$\bN_3(i)$ of the extended universe. Furthermore we have shown that by 
assuming that the 
three-volume fluctuations around $\bN_3(i)$ are sufficiently small
so that a quadratic approximation is valid, we can use the measured fluctuations
to reconstruct the discretized version \rf{n7b} of the minisuperspace action \rf{n5}, 
where $k_1$ and $\tk_2$ are independent of the total four-volume $N_4$ used 
in the simulations. This certainly provides strong evidence that both the
minisuperspace description of the dynamical behaviour of the (expectation value of
the) three-volume, and the semiclassical quadratic truncation for the description of
the quantum fluctuations in the three-volume are essentially correct. 

In the following we will test in more detail how well the actions \rf{n5} and \rf{n7b} 
describe the data encoded in the covariance matrix $\hC$. 
\begin{figure}[t]
\centerline{{\scalebox{1.0}{\rotatebox{0}{\includegraphics{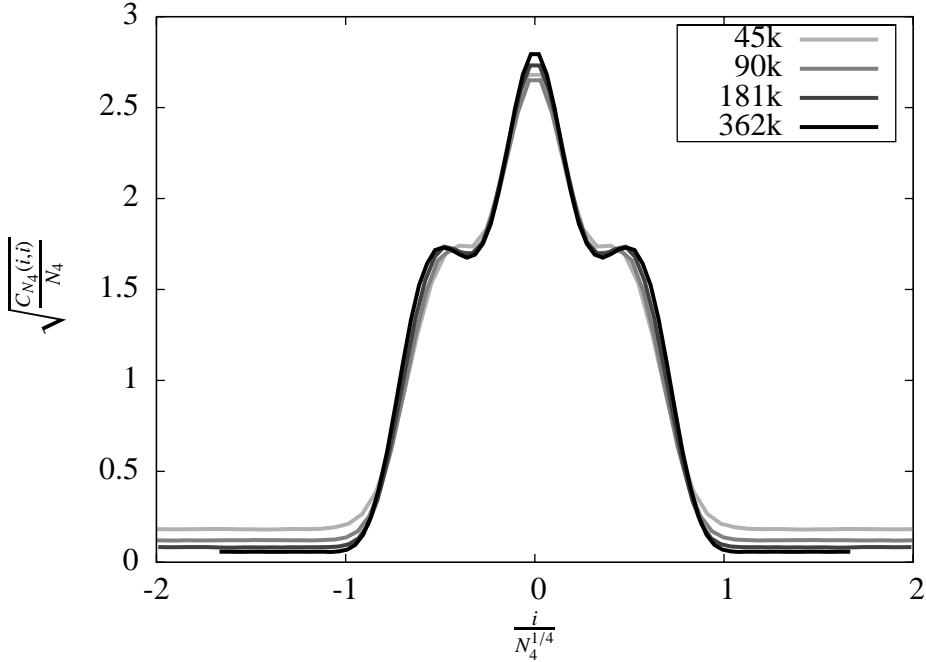}}}}}
\caption{\label{fig7} Analyzing the quantum fluctuations of Fig.\ \ref{fig1}: 
diagonal entries $F(t,t)^{1/2}$ of the universal scaling function 
$F$ from \rf{n7f}, for $N_4^{(4,1)}=$ 20.000,
40.000, 80.000 and 160.000.}
\end{figure}
The correlation function was defined in the previous section by
\beq\label{3h.1}
C_{N_4}(i,i') = \la \delta N_3(i) \delta N_3 (i')\ra,
~~~~\del N_3(i) \equiv N_3(i) -\bN_3(i),
\eeq 
where we have included an additional subscript $N_4$ to emphasize that 
$N_4$ is kept constant in a given simulation. The first observation extracted from
the Monte Carlo simulations is that under a change in the four-volume
$C_{N_4}(i,i')$ scales as\footnote{We stress again that 
the form \rf{n7f} is only valid in that part of the universe whose
spatial extension is considerably 
larger than the minimal $S^3$ constructed from 5 tetrahedra. (The spatial volume
of the stalk typically fluctuates between 5 and 15 tetrahedra.)} 
\beq
C_{N_4}(i,i') =
N_4 \; F\Big({i}/N^{1/4}_4,{i'}/N_4^{1/4}\Big), \label{n7f}
\eeq
where $F$ is a universal scaling function. 
This is illustrated by Fig.\ \ref{fig7} for the rescaled version
of the diagonal part $C_{N_4}^{1/2}(i,i)$, 
corresponding precisely to the quantum fluctuations $\la (\del N_3(i))^2\ra^{1/2}$ 
of Fig.\ \ref{fig1}. While the height of the curve in Fig.\ \ref{fig1}
will grow as $N_4^{3/4}$, the superimposed fluctuations
will only grow as $N_4^{1/2}$. We conclude that {\it for fixed bare 
coupling constants} the relative fluctuations  
will go to zero in the infinite-volume limit . 

From the way the factor $\sqrt{N_4}$ appears as an overall scale in
eq.\ \rf{n7bb} it is clear that to the extent 
a quadratic expansion around the effective background 
geometry is valid one will have a scaling
\beq\label{sc1}
\la \del N_3(i) \del N_3(i')\ra = 
N_4^{3/2} \la \del n_3(t_i) \del n_3(t_{i'})\ra = 
N_4 F(t_i,t_{i'}),
\eeq
where $t_i = i/N_4^{1/4}$. This implies that \rf{n7f} provides additional 
evidence for the validity of the quadratic approximation 
and the fact that our choice of action (\ref{n7b}), with $k_1$ independent of $N_4$
is indeed consistent.

To demonstrate in detail that the full
function $F(t,t')$ and not only its diagonal part is described by the
effective actions \rf{n5}, \rf{n7b}, let us for convenience adopt a 
continuum language and compute its expected behaviour.
Expanding \rf{n5} around the classical solution 
according to $V_3(t) = V_3^{cl}(t) + x(t)$,
the quadratic fluctuations are given by
\bea
\la x(t) x(t')\ra &\! =\! & 
\int \cD x(s)\; x(t)x(t')\; 
e^{ -\frac{1}{2}\int\!\!\int d s d s' x(s) M(s,s') x(s')} \nonumber\\
&\! =\! &  M^{-1}(t,t'),\label{n7a}
\eea
where $\cD x(s)$ is the normalized measure and 
the quadratic form $M(t,t')$ is determined by expanding the 
effective action $S$ to second order in $x(t)$, 
\beq\label{n8}
S(V_3) = S(V_3^{cl}) + \frac{1}{18\pi G}\frac{B}{V_4}  \int \d t \; x(t) \hH 
 x(t).
\eeq
In expression \rf{n8}, $\hH$ denotes the Hermitian operator
\beq\label{n9}
\hH= -\frac{\d }{\d t} \frac{1}{\cos^3 (t/B)}\frac{\d }{\d t} -
\frac{4}{B^2\cos^5(t/B)},
\eeq
which must be diagonalized 
under the constraint that $\int dt \sqrt{g_{tt}}\; x(t) =0$, since
$V_4$ is kept constant. 

Let $e^{(n)}(t)$ be the eigenfunctions of the quadratic form 
given by \rf{n8} with the volume constraint enforced\footnote{One simple
way to find the eigenvalues and eigenfunctions approximately, including
the constraint, is to discretize the differential operator, imposing
that the (discretized) eigenfunctions vanish at the boundaries 
$t= \pm B \pi/2$ and finally adding the constraint as a term $\xi \Big(\int dt\,
x(t)\Big)^2$ to the action, where the coefficient $\xi$ is taken large. The 
differential operator then becomes an ordinary matrix and eigenvalues and
eigenvectors can be found numerically. Stability with respect 
to subdivision and choice of $\xi$ is easily checked.}, 
ordered according to increasing eigenvalues $\l_n$.
As we will discuss shortly, the lowest eigenvalue is $\l_1 =0$,
associated with translational invariance in time direction, and should be left out
when we invert $M(t,t')$, because we precisely fix the centre of volume
when making our measurements. Its dynamics is therefore not accounted for
in the correlator $C(t,t')$.

\begin{figure}[t]
\centerline{{\scalebox{0.8}{\rotatebox{0}{\includegraphics{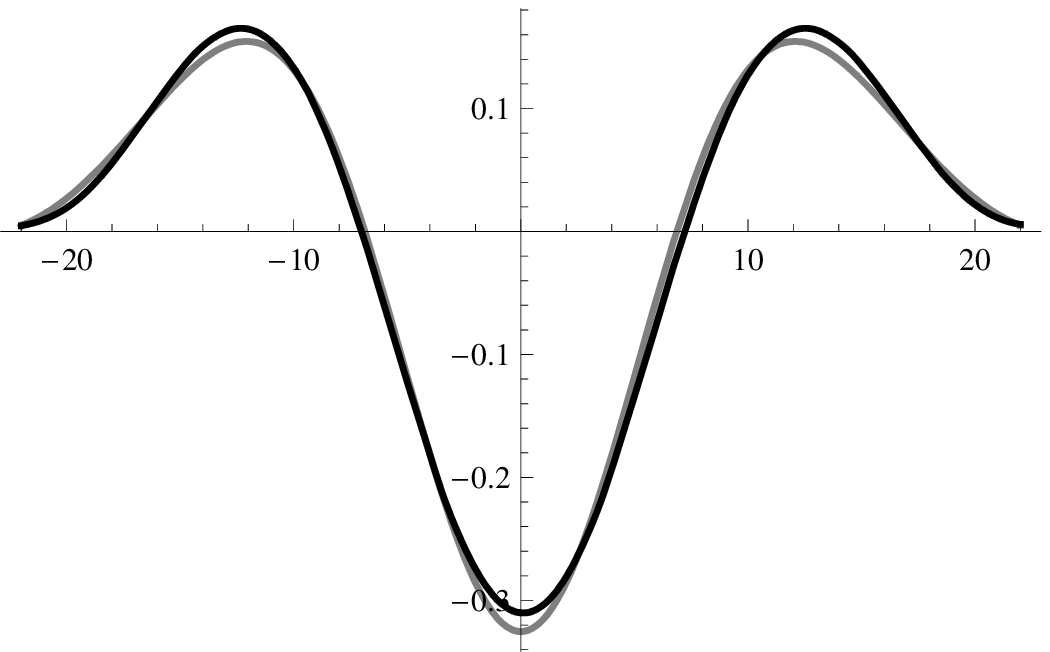}}}}}
\centerline{{\scalebox{0.8}{\rotatebox{0}{\includegraphics{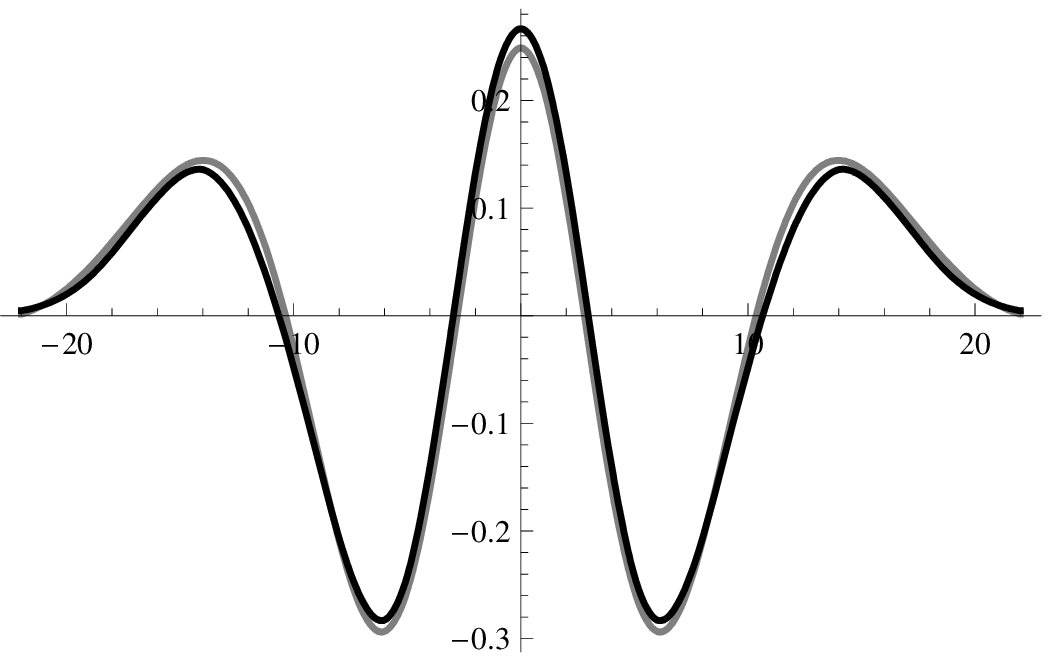}}}}}
\caption{\label{fig8}Comparing the two highest even eigenvector of the 
covariance matrix $C(t,t')$ measured directly
(gray curves) with the two lowest even 
eigenvectors of $M^{-1}(t,t')$, calculated semiclassically (black curves). }
\end{figure}
If this cosmological continuum model were to give the correct 
description of the computer-generated universe, the matrix 
\beq\label{n12}
M^{-1}(t,t') = \sum_{n=2}^\infty \frac{e^{(n)}(t)e^{(n)}(t')}{\l_n}.
\eeq 
should be proportional to the measured correlator $C(t,t')$.
Fig.\ \ref{fig8} shows the eigenfunctions $e^{(2)}(t)$ and $e^{(4)}(t)$
(with two and four zeros respectively),
calculated from $\hH$ with the constraint $\int dt \sqrt{g_{tt}}\; x(t) =0$
imposed. Simultaneously
we show the corresponding eigenfunctions  
calculated from the data, i.e.\ from the matrix $C(t,t')$, which
correspond to the (normalizable) eigenfunctions with the highest and third-highest
eigenvalues. The agreement is very good, 
in particular when taking into consideration that
no parameter has been adjusted in the action (we simply take 
$B=s_0 N_4^{1/4}\Del t$ in \rf{n2} and \rf{n8}, which gives $B=14.47a_t$
for $N_4 = 362.000$).  

The reader may wonder why the first eigenfunction exhibited
has two zeros. As one would expect, the ground state eigenfunction $e^{(0)}(t)$
of the Hamiltonian \rf{n9}, corresponding to the lowest eigenvalue, 
has no zeros, but it does not satisfy the volume
constraint $\int dt \sqrt{g_{tt}}\; x(t) =0$. 
The eigenfunction $e^{(1)}(t)$ of $\hH$ with next-lowest 
eigenvalue has one zero and is given by the simple analytic function
\beq\label{e1}
e^{(1)}(t) = \frac{4}{\sqrt{\pi B}}  \sin \Big(\frac{t}{B}\Big) \, 
\cos^2 \Big(\frac{t}{B}\Big) = c^{-1} \frac{d V_3^{cl} (t)}{dt},
\eeq 
where $c$ is a constant.
One realizes immediately that $e^{(1)}$ is the translational zero mode of the 
classical solution $V_3^{cl}(t)$ ($\propto \cos^3 t/B$). Since the action is 
invariant under time translations we have
\beq\label{cl1}
S(V_3^{cl}(t+\Del t)) = S(V_3^{cl}(t)),
\eeq
and since $V_3^{cl}(t)$ is a solution to the 
classical equations of motion we find to second order (using the 
definition \rf{e1})
\beq\label{cl2}
S(V_3^{cl}(t+\Del t)) = S(V_3^{cl}(t)) + 
\frac{c^2 (\Del t)^2 }{18\pi G}\frac{B}{V_4} \int dt \;e^{(1)}(t) \hH e^{(1)}(t),
\eeq 
consistent with $e^{(1)}(t)$ having eigenvalue zero. 

It is clear from Fig.\ \ref{fig8} that some of the eigenfunctions of $\hH$
(with the volume constraint imposed) agree very well with the measured eigenfunctions.
All even eigenfunctions (those symmetric with respect to reflection about the
symmetry axis located at the centre of volume) turn out to agree very well. 
The odd eigenfunctions of $\hH$ agree less well with the 
eigenfunctions calculated from the measured $C(t,t')$.
The reason seems to be that we have not managed to eliminate the motion
of the centre of volume completely from our measurements. As already mentioned
above, there is an inherent ambiguity in fixing the centre of volume, which 
turns out to be sufficient to reintroduce the zero mode in the data.
Suppose we had by mistake misplaced the centre of volume by a small distance 
$\Del t$.
This would introduce a modification 
\beq\label{mod}
\Del V_3 = \frac{dV_3^{cl}(t)}{dt} \; \Del t
\eeq
proportional to the zero mode of the potential $V_3^{cl}(t)$.
It follows that the zero mode can re-enter whenever we have an ambiguity 
in the position of the centre of volume. In fact, we have found that the first odd eigenfunction
extracted from the data can be perfectly described by 
a linear combination of $e^{(1)}(t)$ and $e^{(3)}(t)$.  
It may be surprising at first that an ambiguity of one lattice
spacing can introduce a significant mixing. However, if we translate 
$\Del V_3$ from eq.\ \rf{mod} to ``discretized'' dimensionless units using 
$V_3(i) \sim N_4^{3/4} \cos (i/N_4^{1/4})$, we find that $\Del V_3 \sim \sqrt{N_4}$,
which because of $\la (\del N_3(i))^2\ra \sim N_4$ is of the same order of magnitude 
as the fluctuations 
themselves. In our case, this
apparently does affect the odd eigenfunctions. 
\begin{figure}[t]
\centerline{{\scalebox{0.6}{\rotatebox{90}{\includegraphics{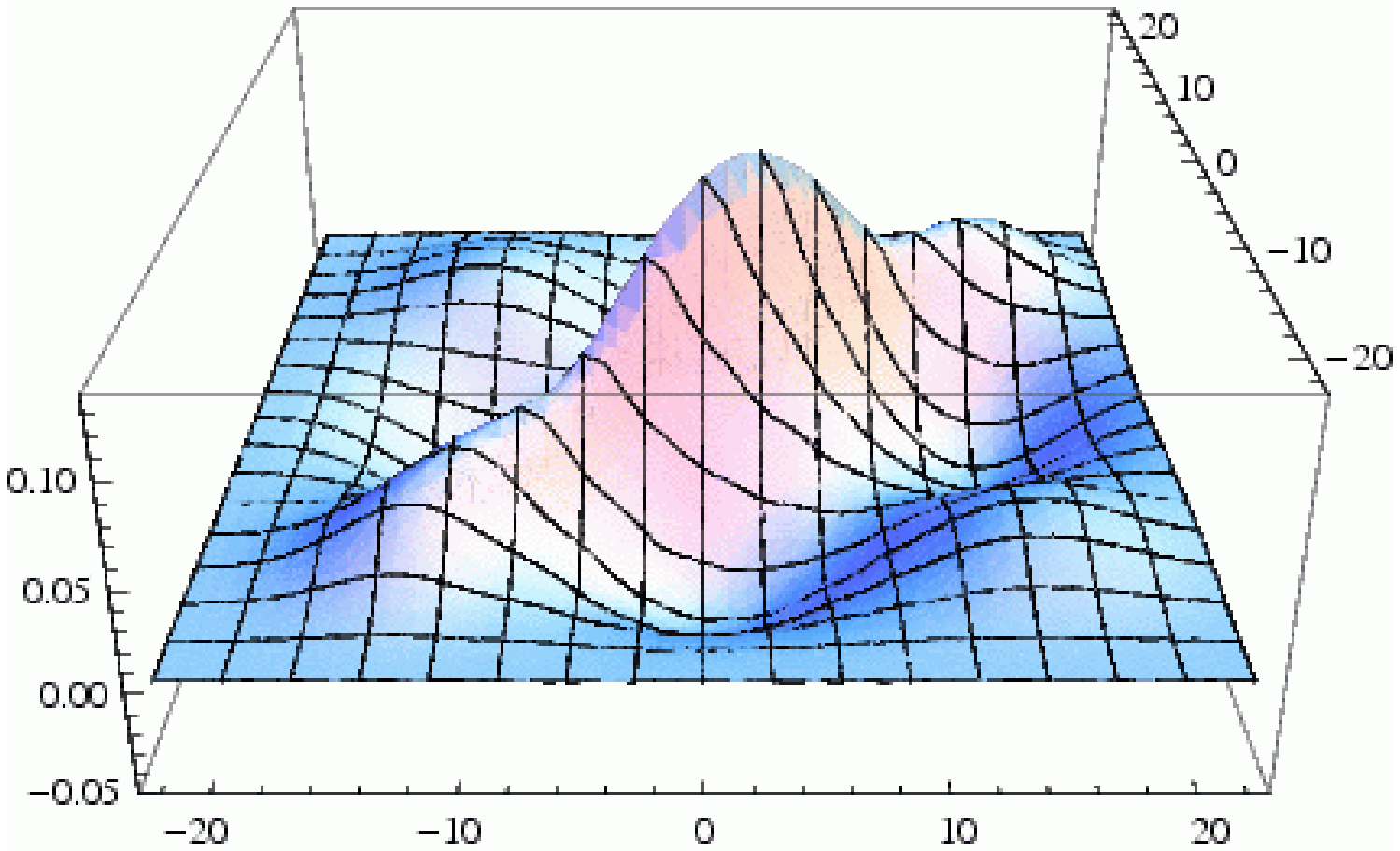}}}}}
\vspace{6pt}
\centerline{{\scalebox{0.6}{\rotatebox{90}{\includegraphics{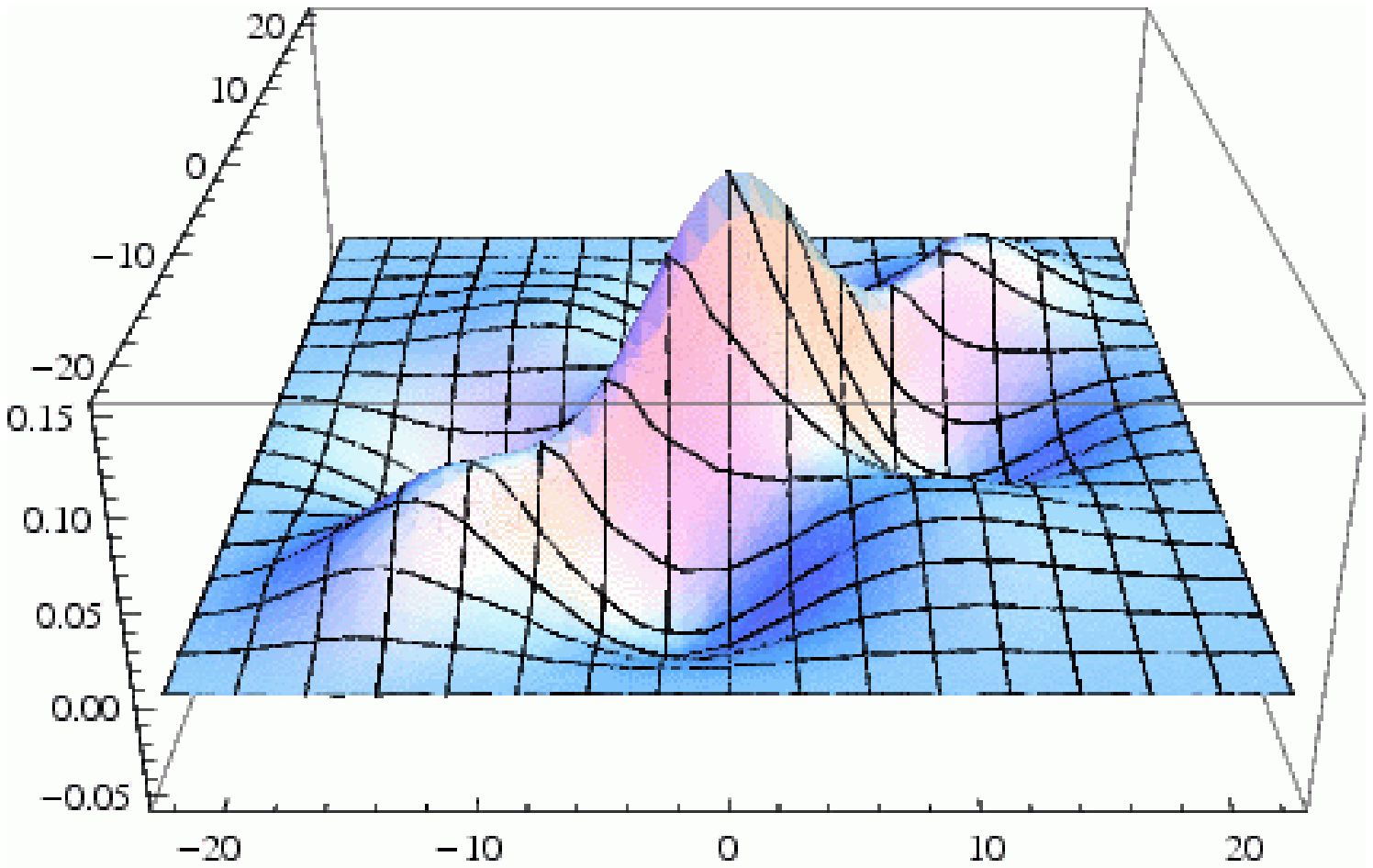}}}}}
\caption{\label{fig9}Comparing data for the extended part of the universe: 
measured $C(t,t')$ (above)
versus $M^{-1}(t,t')$ obtained from analytical calculation (below). 
The agreement is good, and
would have been even better had we included only the even modes.}
\end{figure}

One can also compare the data and 
the matrix $M^{-1}(t,t')$ calculated from \rf{n12} directly. 
This is illustrated in Fig.\ \ref{fig9}, where we have restricted ourselves to
data from inside the extended part of the universe. We imitate the
construction \rf{n12} for $M^{-1}$, using the data to 
calculate the eigenfunctions, rather than $\hH$. One 
could also have used $C(t,t')$ directly, but the use of the 
eigenfunctions makes it somewhat easier to perform the restriction to the bulk.
The agreement is again good (better than 15\% at any point
on the plot), although less spectacular than 
in Fig.\ \ref{fig8} because of the contribution of the odd eigenfunctions to the
data.

\section{The geometry of spatial three-spheres\label{S3}}

We have shown above that our data for the spatial three-volumes 
have a natural interpretation as coming from the slicing of 
a four-sphere with standard geometry (the ``round'' four-sphere), 
with relatively small quantum fluctuations superimposed. It is 
natural to ask to what extent the spatial three-spheres themselves
can be assigned the standard geometry of a ``round" three-sphere, again
with relatively small quantum fluctuations superimposed.
We have already provided evidence that the Hausdorff dimension
of the spatial slices is three \cite{emerge,blp}. 
However, the Hausdorff dimension is 
a very coarse measure of geometry, and even very fractal structures can have
Hausdorff dimension three \footnote{To illustrate the point, the Hausdorff dimension of the 
complex plane (with standard geometry) is of course equal to two, but the same is true
for the highly fractal structure of so-called branched polymers or planar trees embedded 
in the plane.}. 

We have analyzed the geometry of the spatial three-spheres as follows.
Each spatial slice at integer proper time $i$ is a triangulation, consisting of
a certain number $N_3$ of tetrahedra, glued together pairwise such that the 
resulting topology is that of a three-sphere. We now choose an arbitrary tetrahedron as
the origin of measurements and subsequently decompose the $S^3$ into (thick) 
shells of tetrahedra characterized
by their distance $r$ from this origin, where the distance $r$ is defined as the
minimal number of tetrahedra one has to cross when moving from the shell to
the origin via neighbouring tetrahedra. We call the number of tetrahedra
in the shell at distance $r$ the {\it area} $A(r,N_3)$ of the shell.
In order to compute the expectation value of this quantity, we have to
repeat the measurements in a way that averages over different triangulations
of $S^4$, over different spatial slices within the $S^4$'s and over different locations
of the point of origin within those slices. 
In this manner we can test whether 
$\la A(r,N_3)\ra$ behaves like a regular three-sphere (with only small
fluctuations superimposed), with $r$ viewed
as the geodesic distance. If this was the case, one would expect a functional
dependence of the form
\beq\label{7.1}
\la A(r,N_3)\ra \propto N_3^{2/3} \sin^2 \Big( \frac{r}{c N_3^{1/3}}\Big),
\eeq
with $c$ a constant.

\begin{figure}[t]
\centerline{{\scalebox{1.0}{\rotatebox{0}{\includegraphics{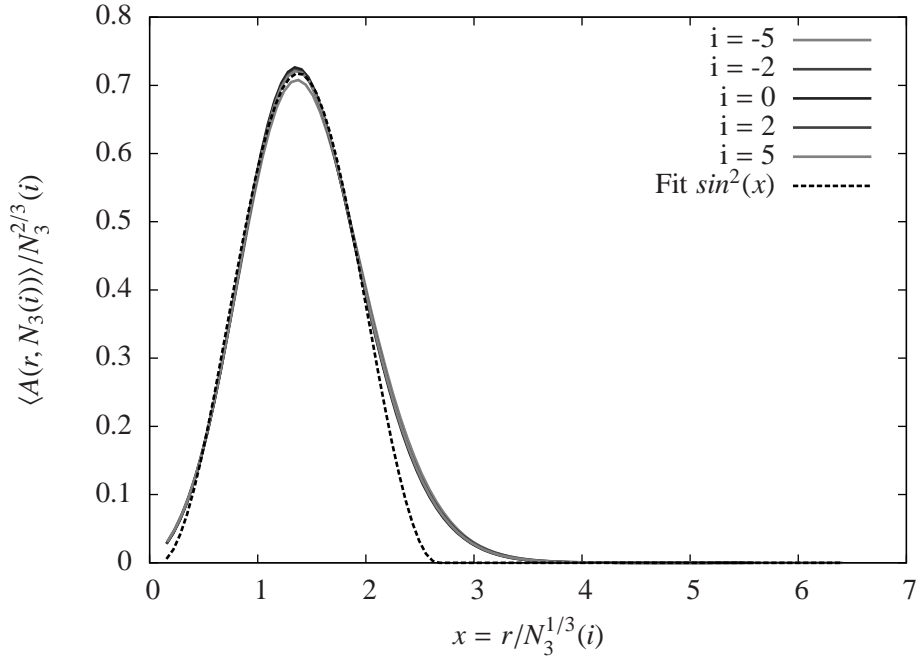}}}}}
\caption{\label{fig10}Testing relation \rf{7.1} for the bare coupling
constants $\k_0 = 2.2$ and  $\Del=0.6$, at four-volume 
$N_4^{(4,1)} = 160.000$. Data have been collected for spatial slices at various
distances close to the centre of volume.}
\end{figure} 

Fig.\ \ref{fig10} summarizes the results of our measurements.
Since we are not interested in very small $N_3$'s where no continuum 
scaling is expected, we have restricted ourselves to spatial slices
close to the centre of volume as defined above, where $N_3$
is largest. The first thing to note about Fig.\ \ref{fig10} is that the data
from spatial slices at different distances from the centre of volume fall to
good accuracy on a common, universal curve. Next, we observe 
that relation \rf{7.1} is reasonably well satisfied, except for the measurements 
at large radii $r$, which exhibit a tail not described by formula
\rf{7.1}. This signals the presence of large fluctuations 
in the geometry (the shape) of the spatial slices to the effect that we cannot 
simply view them -- in the sense of expectation values -- as classical spheres of
constant positive curvature with fixed radius proportional to
$N^{1/3}$, superimposed by small quantum fluctuations.   
In fact there is already evidence that 
the geometry, when defined with respect to the geodesic distance $r$, has
certain fractal properties \cite{blp}. This can be substantiated and quantified by  
measuring the topology of a typical spherical shell
at a distance $r$ from a chosen origin in more detail. At sufficiently large radius $r$ one
finds that the topology is no longer that of a single two-sphere, but branches out
into a number of disconnected pieces, most likely
by effectively creating a number of spatial ``baby universes''.
It is well known how to study the distribution of such
baby universes \cite{ajjk,ajt}, and we believe that these methods will yield
a quantitative description of the observed slower fall-off for large $r$. Details 
of this picture, including a study of the temporal dynamics of
such spatial baby universes, will be published elsewhere.

\section{The size of the universe and the flow of $G$\label{size}}

Let us now return to equation \rf{n7c},
\beq\label{n7cc}
{G} = \frac{a^2}{k_1} \frac{\sqrt{\tC_4}\; \ts_0^2}{3\sqrt{6} },
\eeq
which relates the parameter $k_1$ extracted from the Monte Carlo simulations to 
Newton's constant in units of the cut-off $a$, $G/a^2$. For the bare coupling constants
$(\k_0,\Del)= (2.2,0.6)$ we have high-statistics measurements
for $N_4$ ranging from 45.500 to 362.000 four-simplices (equivalently, $N_4^{(4,1)}$
ranging from 20.000 to 160.000 four-simplices). The choice of 
$\Del$ determines the asymmetry parameter $\a$, and the 
choice of $(\k_0,\Del)$ determines the ratio $\xi$ 
between $N_4^{(3,2)}$ and $N_4^{(4,1)}$. This in turn determines 
the ``effective'' four-volume $\tC_4$ of an average four-simplex, which
also appears in \rf{n7cc}. The number $\ts_0$ in \rf{n7cc} 
is determined directly from the time extension $T_{\rm univ}$ of the extended universe
according to
\beq\label{width}
T_{\rm univ}=\pi\; \ts_0 \Big(N_4^{(4,1)}\Big)^{1/4}.
\eeq
Finally, from our measurements we have determined $k_1= 0.038$. 
Taking everything together according to  \rf{n7cc}, we obtain $G\approx 0.23 a^2$, or 
$\ell_{Pl}\approx 0.48 a$, where  $\ell_{Pl} = \sqrt{G}$ is the Planck length.

From the identification of the volume of the four-sphere, 
$V_4 =  8\pi^2 R^4/{3} = \tC_4 N_4^{(4,1)} a^4$,
we obtain that $R=3.1 a$. In other words, 
{\it the linear size $\pi R$ of the quantum de Sitter universes 
studied here lies in the range of 12-21 Planck lengths for $N_4$ in the 
range mentioned above and for the bare 
coupling constants chosen as $(\k_0,\Del)=(2.2,0.6)$}.\footnote{Small deviations
from the corresponding numbers quoted in \cite{agjl} have their origin in the more
careful (and correct) treatment of the various four-volumes $N_4$, $N_4^{(4,1)}$ and
$N_4^{(3,2)}$ in the present work.}

Our dynamically generated universes are therefore not very big, and the 
quantum fluctuations around their average shape are large as is apparent from 
Fig.\ \ref{fig1}. It is rather 
surprising that the semiclassical minisuperspace formulation is applicable
for universes of such a small size, a fact that should be welcome news to anyone
performing semiclassical calculations to describe the behaviour of the early universe.
However, in a certain sense our lattices are still coarse compared
to the Planck scale $\ell_{Pl}$ because the Planck length is roughly half a lattice
spacing. If we are after a theory of quantum gravity valid on all scales, 
we are in particular interested in uncovering phenomena associated with Planck-scale
physics. In order to collect data free from unphysical short-distance lattice artifacts at this
scale, we would ideally like to work with a
lattice spacing much smaller than the Planck length,
while still being able to set by hand the physical volume of the  
universe studied on the computer.

The way to achieve this, under the assumption that the coupling 
constant  $G$ of formula \rf{n7cc} is indeed a true measure of the gravitational
coupling constant, is as follows.
We are free to vary the discrete four-volume $N_4$ and the bare coupling 
constants $(\k_0, \Del)$ of the Regge action (see \cite{blp}
for further details on the latter). Assuming for the
moment that the semiclassical minisuperspace action is valid, 
the effective coupling constant $k_1$ in front of it
will be a function of the bare coupling constants $(\k_0, \Del)$,
and can in principle be determined as described above for the case 
$(\k_0, \Del)=(2.2,0.6)$. If we adjusted the bare coupling 
constants such that in the limit as $N_4 \to \infty$ both
\beq\label{n100}
V_4 \sim N_4 a^4~~~{\rm and}~~~G\sim a^2/k_1(\k_0,\Del)
\eeq
remained constant (i.e.\ $k_1(\k_0,\Del) \sim 1/\sqrt{N_4}$), 
we would eventually reach a region where the Planck length was 
significantly smaller than the lattice spacing $a$, in which event the lattice
could be used to approximate spacetime structures of Planckian size
and we could initiate a genuine study of the sub-Planckian regime.
\begin{figure}[th]
\centerline{{\scalebox{0.62}{\rotatebox{0}{\includegraphics{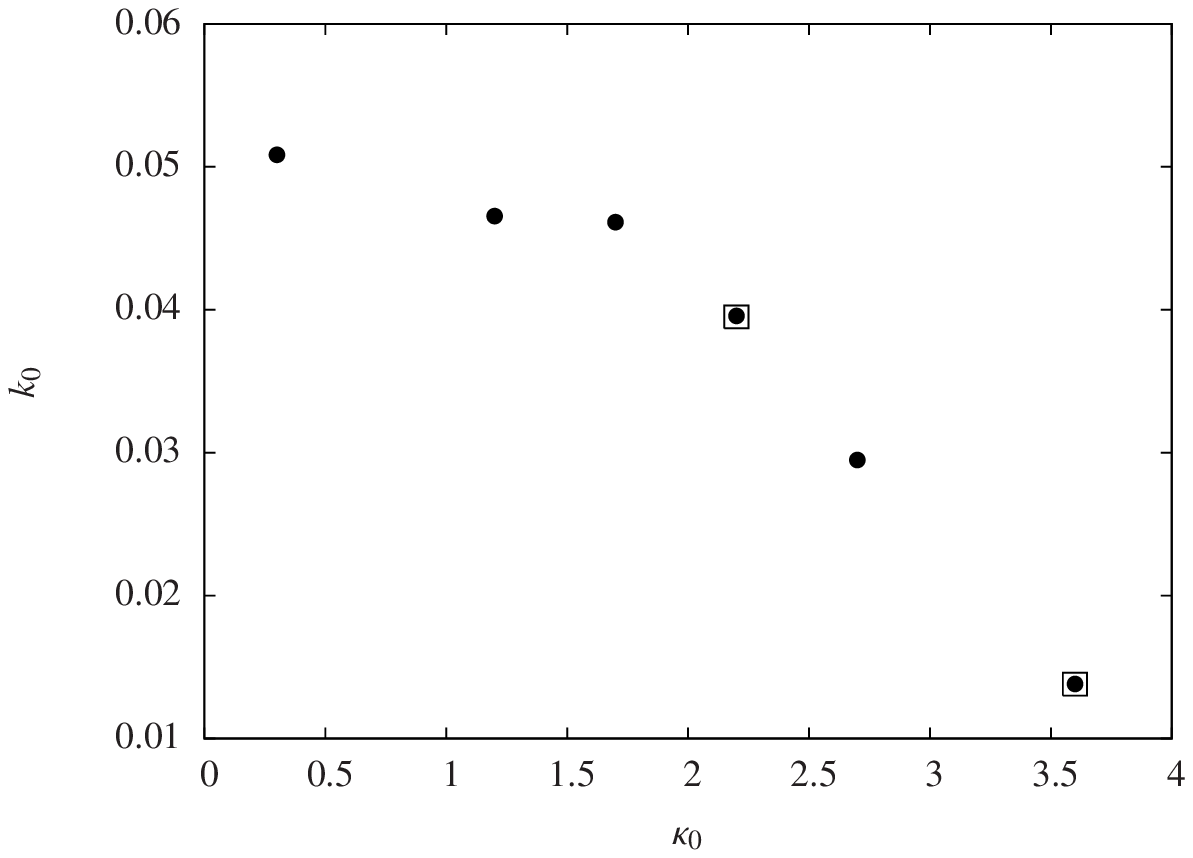}}}}}
\centerline{{\scalebox{0.62}{\rotatebox{0}{\includegraphics{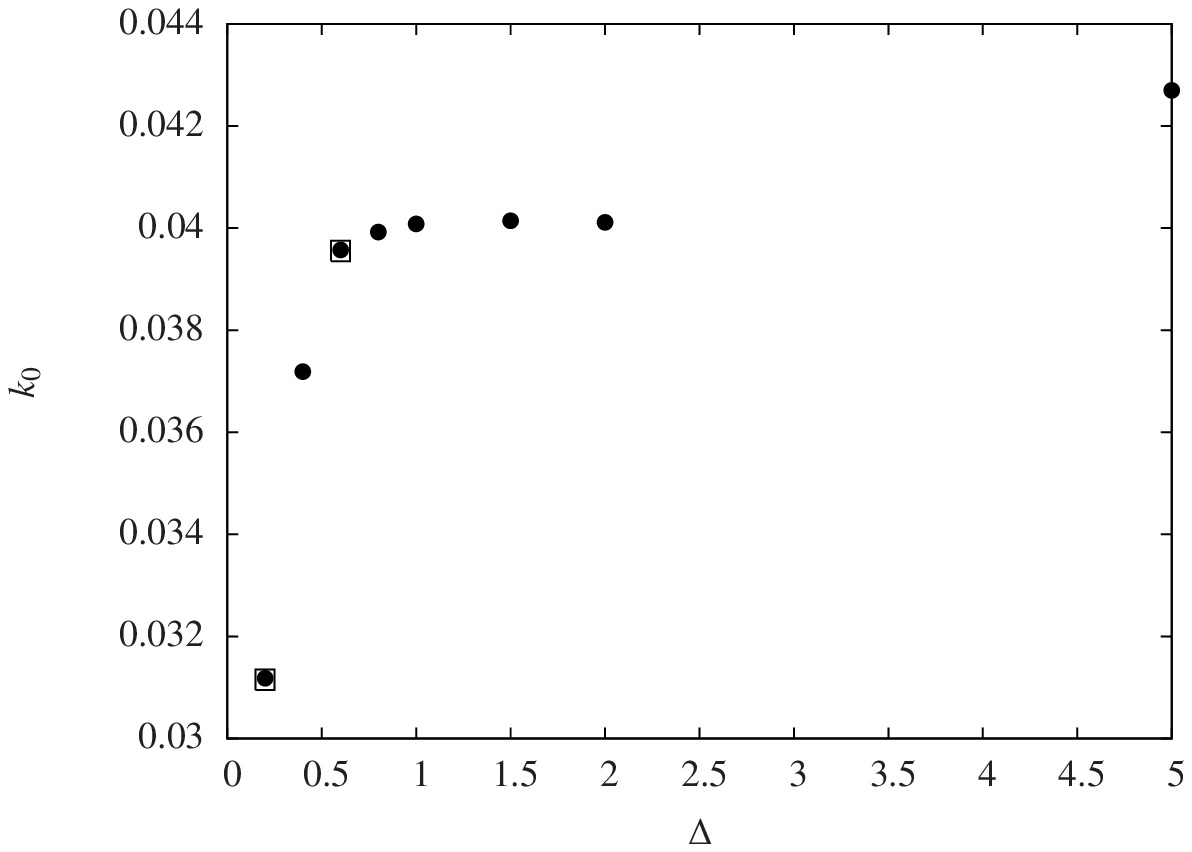}}}}}
\caption{\label{fig12}The measured effective 
coupling constant $k_1$ as function
of the bare $\k_0$ (top, $\Del = 0.6$ fixed) and the asymmetry $\Del$ (bottom, 
$\k_0=2.2$ fixed). The marked point near the middle of the data points sampled
is the point $(\k_0,\Del)=(2.2,0.6)$ where most measurements in the remainder of the
paper were taken. The other marked points are those closest to the two phase 
transitions, to the ``branched-polymer phase" (top), and the ``crumpled phase" (bottom).}
\end{figure} 
Since we have no control over the effective coupling constant 
$k_1$, the first obvious question which arises is whether we can at all
adjust the bare coupling constants in such a way that at large 
scales we still see a four-dimensional universe, with
$k_1$ going to zero at the same time. The answer seems to be in the affirmative,
as we will go on to explain.

Fig.\ \ref{fig12} shows the results of extracting 
$k_1$ for a range of bare coupling constants for which we still
observe an extended universe. In the top figure 
$\Del = 0.6$ is kept constant while $\k_0$ is varied. For $\k_0$ sufficiently large
we eventually reach a point where a phase transition takes place (the point
in the square in the bottom right-hand corner is the measurement closest to
the transition we have looked at).    
For even larger values of $\k_0$, beyond this transition, the universe disintegrates 
into a number of small universes, in a CDT-analogue of the branched-polymer phase of
Euclidean quantum gravity. The plot shows that 
the effective coupling constant $k_1$ becomes smaller and 
possibly goes to zero as the phase transition point is approached, although
our current data do not yet allow us to conclude that $k_1$ 
does indeed vanish at the transition point.

Conversely, the bottom figure of Fig.\ \ref{fig12} shows the effect of varying  
$\Del$, while keeping $\k_0=2.2$ fixed.
As $\Del$ is decreased towards 0, we 
eventually hit another phase transition, separating the physical phase of 
extended universes from the CDT-equivalent of the 
crumpled phase of Euclidean quantum gravity, where the 
entire universe will be concentrated within a few time steps,
as already mentioned in Sec.\ \ref{S4} above. (The point closest to the transition where
we have taken measurements is the one in the bottom left-hand corner.)
Also when approaching this phase transition
the effective coupling constant $k_1$ goes to 0, leading to the tentative 
conclusion that $k_1 \to 0$ along the entire phase boundary.

However, to extract the coupling constant $G$ 
from \rf{n7cc} we not only have to take into account the change in $k_1$, but also
that in $\ts_0$ (the width of the distribution $N_3(i)$) and 
in the effective four-volume $\tC_4$ as a 
function of the bare coupling constants.
Combining these changes, we arrive at
a slightly different picture. Approaching the boundary where spacetime
collapses in time direction (by lowering $\Del$), the
gravitational coupling constant $G$ {\it decreases},
despite the fact that $1/k_1$ increases. This is a consequence of $\ts_0$ 
decreasing considerably, as can be seen from Fig.\ \ref{fig3}. 
On the other hand, when (by increasing $\k_0$) we approach the 
region where the universe breaks up into several independent components, 
the effective gravitational coupling constant $G$ increases, more or 
less like $1/k_1$, where the behaviour of $k_1$ is shown in 
Fig.\ \ref{fig12} (top). This implies that  
the Planck length $\ell_{Pl} = \sqrt{G}$ increases 
from approximately $0.48 a$ to $0.83 a$ when $\k_0$ changes from
2.2 to 3.6. Most likely we can make it even bigger in terms of Planck units
by moving closer to the phase boundary.       

On the basis of these arguments, it seems likely that the nonperturbative
CDT-formulation of quantum gravity does allow us to 
penetrate into the sub-Planckian regime and probe the physics there explicitly. 
Work in this direction is currently ongoing. 
One interesting issue under investigation is whether and to what extent 
the simple minisuperspace description remains valid as we go to shorter scales.
We have already seen deviations from classicality at short scales
when measuring the spectral dimension \cite{spectral,blp}, and one would expect
them to be related to additional terms in the 
effective action \rf{n5} and/or a nontrivial scaling behaviour of $k_1$.
This raises the interesting possibility of being able to test explicitly the
scaling violations of $G$ predicted by renormalization group methods
in the context of asymptotic safety \cite{reuteretc}.

\section{Discussion\label{discussion}}

The CDT model of quantum gravity is extremely simple. It is the path integral 
over the class of causal geometries with a global time foliation. In order
to perform the summation explicitly, we introduce a grid of piecewise linear 
geometries, much in the same way as when defining the path integral in 
quantum mechanics. Next, we rotate each of these geometries to Euclidean 
signature and use as bare action the Einstein-Hilbert action\footnote{Of course, 
the full, effective action, including measure contributions, will contain all higher-derivative 
terms.} in Regge form. That is all.

The resulting superposition exhibits a nontrivial scaling behaviour as function of the 
four-volume, and we observe the appearance of a well-defined average 
geometry, that of de Sitter space, the maximally symmetric solution to the classical
Einstein equations in the presence of a positive cosmological constant.  
We are definitely in a quantum regime,
since the fluctuations of the three-volume around de Sitter space are sizeable, 
as can be seen in 
Fig.\ \ref{fig1}. Both the average geometry and the quantum fluctuations 
are well described in terms of the mini\-superspace action \rf{n5}. 
A key feature to appreciate is that, unlike in standard (quantum-)cosmological treatments, 
this description is the {\it outcome} of a nonperturbative evaluation
of the {\it full} path integral, 
with everything but the scale factor (equivalently, $V_3(t)$)
summed over. Measuring the correlations of the quantum fluctuations in the computer
simulations for a particular choice of bare coupling constants
enabled us to determine the continuum gravitational coupling
constant $G$ as $G\approx 0.42 a^2$, thereby introducing an absolute physical 
length scale into the dimensionless lattice setting. 
Within measuring accuracy, our de Sitter universes (with volumes lying
in the range of 6.000-47.000 $\ell_{Pl}^4$) are seen to behave perfectly 
semiclassically with regard to their large-scale properties. 

We have also indicated how we may be able to penetrate  
into the sub-Planckian regime by suitably changing the bare coupling constants.
By ``sub-Planckian regime" we mean that  
the lattice spacing $a$ is (much) smaller than 
the Planck length. While we have not yet analyzed this region in
detail, we expect to eventually observe a breakdown of the 
semiclassical approximation.
This will hopefully allow us to make contact with
attempts to use renormalization group techniques in the continuum
and the concept of asymptotic safety 
to study scaling violations in quantum gravity \cite{reuteretc}.

On the basis of the results presented here, two major issues suggest 
themselves for further research. First, we need to establish  
the relation of our effective gravitational coupling constant $G$ 
with a more conventional gravitational 
coupling constant, defined directly in terms of coupling 
matter to gravity. In the present work, we have defined $G$ as the coupling
constant in front of the effective action, but it would be desirable to verify
directly that a gravitational coupling defined via the coupling to matter agrees
with our $G$. In principle it is easy to couple matter to 
our model, but it is less straightforward to define in a simple way 
a set-up for extracting the semiclassical effect of gravity on the matter sector. 
Attempts in this direction were already undertaken in the ``old'' Euclidean 
approach \cite{js,newton}, and it is possible that similar ideas can be used
in CDT quantum gravity. Work on this is in progress.

The second issue concerns the precise nature 
of the ``continuum limit''. Recall our discussion in the Introduction
about this in a conventional lattice-theoretic setting. The 
continuum limit is usually linked to a divergent correlation length
at a critical point. It is unclear whether such a scenario is realized in our case.
In general, it is rather unclear how one could define at all 
the concept of a divergent length related to correlators in quantum gravity, 
since one is integrating over all geometries, and it is the geometries
which dynamically give rise to the notion of ``length".

This has been studied in detail in two-dimensional (Euclidean) 
quantum gravity coupled to matter with central charge $c \le 1$ 
\cite{corr2d}. It led to the conclusion that one
could associate the critical behaviour of the 
matter fields (i.e.\ approaching the critical point of 
the Ising model) with a divergent correlation length, although 
the matter correlators themselves had to be defined
as non-local objects due to the requirement
of diffeomorphism invariance. On the other hand, the two-dimensional
studies do not give us a clue of how to treat the 
gravitational sector itself, since they do not possess 
gravitational field-theoretic degrees of freedom. 
What happens in the two-dimensional lattice models
which can be solved analytically is that the only fine-tuning needed
to approach the continuum limit is an additive renormalization
of the cosmological constant (for fixed matter couplings).
Thus, fixing the two-dimensional spacetime volume $N_2$ (the 
number of triangles), such that the cosmological constant plays no role,
there are no further coupling constants to adjust and the continuum limit is 
automatically obtained by the assignment $V_2 = N_2 a^2$ and
taking $N_2 \to \infty$. This situation can also occur in special 
circumstances in ordinary lattice field theory. A term like
\beq\label{k5}
\sum_i c_1 (\phi_{i+1}-\phi_i)^2 + c_2(\phi_{i+1}+\phi_{i-1}-2\phi_i)^2
\eeq
(or a higher-dimensional generalization) will also go to the continuum free 
field theory simply by increasing the lattice size and using
the identification $V_d = L^d a^d$ ($L$ denoting the linear size of 
the lattice in lattice units), the higher-derivative term being 
sub-dominant in the limit. It is not obvious that in quantum gravity 
one can obtain a continuum quantum field theory without fine-tuning in
a similar way, because the action in this case is multiplied by  
a dimensionful coupling constant. Nevertheless, it is certainly remarkable
that the infrared limit of our effective action apparently 
reproduces -- within the cosmological setting -- the Einstein-Hilbert action, which is
the unique diffeomorphism-invariant generalization of the 
ordinary kinetic term, containing at most second derivatives
of the metric. A major question is whether and how far our theory can 
be pushed towards an ultraviolet limit. We have indicated how to obtain such 
a limit by varying the bare coupling constants of the theory, but 
the investigation of the limit $a \to 0$ with fixed $G$ has only just begun.

\subsection*{Acknowledgments}
All authors acknowledge support by
ENRAGE (European Network on
Random Geometry), a Marie Curie Research Training Network, 
contract MRTN-CT-2004-005616, and 
AG and JJ by COCOS (Correlations in Complex Systems), 
a Marie Curie Transfer
of Knowledge Project, contract MTKD-CT-2004-517186, both in the 
European Community's Sixth Framework Programme.
RL acknowledges support by the Netherlands
Organisation for Scientific Research (NWO) under their VICI
program.
JJ acknowledges a partial support by the Polish Ministry of Science and
Information Technologies grant 1P03B04029 (2005-2008)


\begin{thebibliography}{99}

\bibitem{weinberg}
S.~Weinberg:
{\it Ultraviolet divergences in quantum theories of gravitation},
in {\it General relativity: Einstein centenary survey}, eds. S.W.\ Hawking
and W.\ Israel, Cambridge University Press, Cambridge, UK (1979) 790-831.

\bibitem{reuteretc}
A.~Codello, R.~Percacci and C.~Rahmede:
{\it Investigating the ultraviolet properties of gravity with a Wilsonian
renormalization group equation}
[0805.2909, hep-th];\\
M.~Reuter and F.~Saueressig:
{\it Functional renormalization group equations, asymptotic safety, and Quantum
Einstein Gravity} [0708.1317, hep-th];\\
M.~Niedermaier and M.~Reuter:
{\it The asymptotic safety scenario in quantum gravity},
Living Rev.\ Rel.\ 9 (2006) 5;\\
H.W.~Hamber and R.M.~Williams:
{\it Nonlocal effective gravitational field equations and the running of
Newton's G},
Phys.\ Rev.\  D\ 72 (2005) 044026 [hep-th/0507017];\\
D.F.~Litim:
{\it Fixed points of quantum gravity},
Phys.\ Rev.\ Lett.\ 92 (2004) 201301 [hep-th/0312114];\\
H.~Kawai, Y.~Kitazawa and M.~Ninomiya:
{\it Renormalizability of quantum gravity near two dimensions,}
Nucl.\ Phys.\ B\ 467 (1996) 313-331 [hep-th/9511217].

\bibitem{ajl4d}
J.~Ambj\o rn, J.~Jurkiewicz and R.~Loll:
{\it Dynamically triangulating Lorentzian quantum gravity,}
Nucl.\ Phys.\ B 610 (2001) 347-382 [hep-th/0105267].

\bibitem{blp}
J.\ Ambj\o rn, J.\ Jurkiewicz and R.\ Loll: 
{\it Reconstructing the universe},
Phys.\ Rev.\ D\ 72 (2005) 064014  [hep-th/0505154].

\bibitem{al} J.\ Ambj\o rn and R.\ Loll:
{\it Non-perturbative Lorentzian quantum gravity, causality and
topology change},
Nucl.\ Phys.\ B\ 536 (1998) 407-434 [hep-th/9805108].

\bibitem{emerge}
J.~Ambj\o rn, J.~Jurkiewicz and R.~Loll:
{\it Emergence of a 4D world from causal quantum gravity},
Phys.\ Rev.\ Lett.\ 93 (2004) 131301 [hep-th/0404156].

\bibitem{semi}
J.~Ambj\o rn, J.~Jurkiewicz and R.~Loll:
{\it Semiclassical universe from first principles},
Phys.\ Lett.\ B 607 (2005) 205-213
[hep-th/0411152].

\bibitem{agjl}
J.~Ambj\o rn, A.~G\"orlich, J.~Jurkiewicz and R.~Loll:
{\it Planckian birth of the quantum de Sitter universe},
Phys.\ Rev.\ Lett.\ 100 (2008) 091304 [0712.2485, hep-th].

\bibitem{desitter}
J.~Ambj\o rn, J.~Jurkiewicz and R.~Loll:
{\it The self-organized de Sitter universe}
[0806.0397, gr-qc].

\bibitem{causality}
J.~Ambj\o rn, J.~Jurkiewicz and R.~Loll:
{\it The universe from scratch},
Contemp.\ Phys.\ 47 (2006) 103-117 [hep-th/0509010];\\
R.\ Loll:
{\it The emergence of spacetime, or, quantum gravity on your desktop},
Class.\ Quant.\ Grav.\ 25 (2008) 114006 [0711.0273, gr-qc].
  
\bibitem{teitelboim}
C.~Teitelboim:
{\it Causality versus gauge invariance in quantum gravity and supergravity},
Phys.\ Rev.\ Lett.\ 50 (1983) 705-708;\\
{\it The proper time gauge in quantum theory of gravitation},
Phys.\ Rev.\  D\ 28 (1983) 297-309.

\bibitem{alwz}
J.~Ambj\o rn, R.~Loll, W.~Westra and S.~Zohren:
{\it Putting a cap on causality violations in CDT,}
JHEP {\bf 0712} (2007) 017 [0709.2784, gr-qc];\\
J.~Ambj\o rn, R.~Loll, Y.~Watabiki, W.~Westra and S.~Zohren:
{\it A string field theory based on causal dynamical triangulations,}
JHEP 0805 (2008) 032 [0802.0719, hep-th];\\
{\it A matrix model for 2D quantum gravity defined by causal dynamical
triangulations},
Phys.\ Lett.\  B\ 665 (2008) 252-256 [0804.0252, hep-th].

\bibitem{worm}
J.\ Ambj\o rn, J.\ Jurkiewicz, R.\ Loll and G.\ Vernizzi:
{\it Lorentzian 3d gravity with wormholes via matrix models},
JHEP 0109 (2001) 022 [hep-th/0106082];\\
{\it 3D Lorentzian quantum gravity from the asymmetric ABAB matrix model},
Acta Phys.\ Polon.\ B\ 34 (2003) 4667-4688 [hep-th/0311072];\\
J.~Ambj\o rn, J.~Jurkiewicz and R.~Loll:
{\it Renormalization of 3d quantum gravity from matrix models,}
Phys.\ Lett.\ B\ 581 (2004) 255-262 [hep-th/0307263].

\bibitem{blz} D.~Benedetti, R.~Loll and F.~Zamponi:
{\it (2+1)-dimensional quantum gravity as the continuum limit of causal
dynamical triangulations}, 
Phys.\ Rev.\ D\ 76 (2007) 104022 [0704.3214, hep-th].

\bibitem{adf}
J.~Ambj\o rn, B.~Durhuus and J.~Fr\"ohlich:
{\it Diseases of triangulated random surface models, and possible cures},
Nucl.\ Phys.\  B\ 257 (1985) 433-449;\\
J.~Ambj\o rn, B.~Durhuus, J.~Fr\"ohlich and P.~Orland:
{\it The appearance of critical dimensions in regulated string theories},
Nucl.\ Phys.\  B\ 270 (1986) 457-482.

\bibitem{david}
A.~Billoire and F.~David:
{\it Microcanonical simulations of randomly triangulated planar random
surfaces},
Phys.\ Lett.\  B\ 168 (1986) 279-283.

\bibitem{migdal}
D.V.~Boulatov, V.A.~Kazakov, I.K.~Kostov and A.A.~Migdal:
{\it Analytical and numerical study of the model of dynamically triangulated
random surfaces},
Nucl.\ Phys.\  B\ 275 (1986) 641-686.

\bibitem{aj}
J.~Ambj\o rn and J.~Jurkiewicz:
{\it Four-dimensional simplicial quantum gravity},
Phys.\ Lett.\  B\ 278 (1992) 42-50.

\bibitem{migdal1}
M.E.~Agishtein and A.A.~Migdal,
{\it Simulations of four-dimensional simplicial quantum gravity},
Mod.\ Phys.\ Lett.\  A\ 7 (1992) 1039-1062.

\bibitem{spectral}
J.~Ambj\o rn, J.~Jurkiewicz and R.~Loll:
{\it Spectral dimension of the universe}, 
Phys.\ Rev.\ Lett.\ 95 (2005) 171301
[hep-th/0505113].

\bibitem{dl} B. Dittrich and R. Loll:
{\it Counting a black hole in Lorentzian product triangulations},
Class.\ Quant.\ Grav. 23 (2006) 3849-3878 [gr-qc/0506035].

\bibitem{js}
B.V.~de Bakker and J.~Smit:
{\it Gravitational binding in 4D dynamical triangulation},
Nucl.\ Phys.\  B\ 484 (1997) 476-494 [hep-lat/9604023].

\bibitem{newton}
H.W.~Hamber and R.M.~Williams:
{\it Newtonian potential in quantum Regge gravity},
Nucl.\ Phys.\ B\ 435 (1995) 361-398 [hep-th/9406163].

\bibitem{corr2d}       
J.~Ambj\o rn, K.N.~Anagnostopoulos, U.~Magnea and G.~Thorleifsson:
{\it Geometrical interpretation of the KPZ exponents},
Phys.\ Lett.\  B\ 388 (1996) 713-719 [hep-lat/9606012];\\
J.~Ambj\o rn and K.N.~Anagnostopoulos:
{\it Quantum geometry of 2D gravity coupled to unitary matter},
Nucl.\ Phys.\  B\ 497 (1997) 445-478 [hep-lat/9701006];\\
J.~Ambj\o rn, K.~N.~Anagnostopoulos and R.~Loll:
{\it A new perspective on matter coupling in 2d quantum gravity},
Phys.\ Rev.\  D\ 60 (1999) 104035 [hep-th/9904012].

\bibitem{ajjk}
J.~Ambj\o rn, S.~Jain, J.~Jurkiewicz and C.F.~Kristjansen:
{\it Observing 4-d baby universes in quantum gravity},
Phys.\ Lett.\  B\ 305 (1993) 208 [hep-th/9303041].

\bibitem{ajt}
J.~Ambj\o rn, S.~Jain and G.~Thorleifsson:
{\it Baby universes in 2-d quantum gravity},
Phys.\ Lett.\  B\ 307 (1993) 34-39 [hep-th/9303149].



\end{thebibliography}
\end{document}